\documentclass[prd,english,nofootinbib,preprintnumbers]{revtex4}
\usepackage[T1]{fontenc}
\usepackage[latin1]{inputenc}
\usepackage{amsmath}
\usepackage{graphicx}
\usepackage{amssymb}

\def\sf{\rm }

\makeatletter

\newcommand{\noun}[1]{\textsc{#1}}

\usepackage{bm}
\newcommand{\geff}{g_{\rm eff}}
\newcommand{\vac}{{\rm vac}}
\newcommand{\muMS}{\bar\mu_{\overline{\rm MS}}}
\newcommand{\mubar}{\bar\mu}

\newcommand{\im}{{\rm Im}}
\newcommand{\re}{{\rm Re}}

\newcommand{\omegaplhat}{{\hat \omega}_{\rm pl.}}

\newcommand{\Li}{{\rm Li}}

\def\q{{\bm q}}

\def\Im{\,{\rm Im}\,}
\newcommand{\be}{\begin{equation}}
\newcommand{\ee}{\end{equation}}
\newcommand{\bea}{\begin{eqnarray}}
\newcommand{\eea}{\end{eqnarray}}

\newcommand{\0}{\over }
\newcommand{\2}{{1\over2}}
\newcommand{\4}{{1\over4}}

\def\0{\over } \def\2{{1\over2}} \def\4{{1\over4}}
\def\5{\hat } \def\6{\partial }

\def\g{g_{\rm eff}}

\usepackage{babel}
\makeatother
\begin{document}

\preprint{ECT*-05-07}
\preprint{TUW-05-11}

\title{Study of the gluon propagator in the large-$N_{f}$ limit at finite
temperature\\ and chemical potential for weak and strong couplings}

\author{Jean-Paul Blaizot}

\email{blaizot@ect.it}

\affiliation{ECT{*}, Villa Tambosi, Strada delle Tabarelle 286,\\ I-38050 Villazzano
Trento, ITALY}

\author{Andreas Ipp}

\email{ipp@ect.it}

\affiliation{ECT{*}, Villa Tambosi, Strada delle Tabarelle 286,\\ I-38050 Villazzano
Trento, ITALY}

\author{Anton Rebhan}

\email{rebhana@hep.itp.tuwien.ac.at}

\affiliation{Institut f$\ddot{u}$r Theoretische Physik, Technische Universit$\ddot{a}$t
Wien\\ Wiedner Hauptstrasse 8-10/136, A-1040 Wien, AUSTRIA.}

\date{\today}

\begin{abstract}
At finite temperature and chemical potential, the leading-order
(hard-thermal-loop) contributions to the gauge-boson
propagator lead to momentum-dependent
thermal masses for propagating quasiparticles as well
as dynamical screening and Landau damping effects. We
compare the
hard-thermal-loop propagator with the complete
large-$N_{f}$ gluon propagator, for which the
usually subleading contributions, such as a finite width of
quasiparticles, can be studied at nonperturbatively
large effective coupling. {\sf We also study quantitatively
the effect of Friedel oscillations in low-temperature electrostatic screening.}
\end{abstract}
\maketitle

\section{Introduction}

At finite temperature and chemical potential, propagators of elementary
fields are modified importantly at soft momentum scales. At leading
order, scalar fields
acquire simple thermal mass terms, but fermions and gauge fields
develop additional quasiparticle branches with complicated
momentum dependent thermal masses for propagating modes,
Landau damping cuts, and, in the case of gauge fields,  
poles for imaginary wave vectors corresponding to dynamical screening
\cite{Silin:1960,Kalashnikov:1980cy,Klimov:1981ka,Weldon:1982aq,Weldon:1989ys,Pisarski:1989wb}.

While the physical singularities of propagators in a gauge theory
are gauge-fixing independent \cite{Kobes:1991dc},
naive perturbation theory leads
to gauge-fixing dependent results beyond leading order
and requires (at least) resummation of the full
nonlinear and nonlocal hard-thermal-loop (HTL) effective action
\cite{Braaten:1990mz} (see e.g.\ Ref.~\cite{Kraemmer:2003gd} for
a recent review).
It has also been found that thermodynamic quantities like the entropy of QCD
can be understood well down to temperatures $T\gtrsim3T_{c}$
with couplings as large as $g\approx2$ by using
HTL quasiparticles in $\Phi$-derivable two-loop approximations \cite{Blaizot:2003tw} (see also \cite{Andersen:2004fp}).
One would therefore want to understand the underlying quasi-particle
picture also at larger couplings. The success of those resummation
techniques has been established so far only by comparison to lattice
data, therefore an independent test is desirable. One possibility
to go beyond HTL is the large flavor number ($N_{f}$) limit \cite{Moore:2002md,Ipp:2003zr,Ipp:2003jy}
in which thermodynamic quantities like pressure or entropy can be
calculated to all orders in the effective coupling, at next-to-leading
order of the $1/N_{f}$ expansion (which is of order $N_f^0$).
In the large-$N_f$ limit, the thermodynamic potential exhibits
a non-trivial (non-monotonic)
behavior when going from weak to strong coupling.
As will be shown in a forthcoming paper \cite{Birr}, the latter is
due to a corresponding nontrivial behavior of the next-to-leading
order asymptotic mass of the quarks. However, the contributions
of the gauge bosons, which are needed only to
leading order in the large-$N_f$ expansion, are of similar magnitude,
and it is worth to study this propagator and its properties
in detail, in particular how it compares with the corresponding
HTL propagator.

The large-$N_f$ gluon propagator is essentially abelian and one-loop, and
some comparisons with the HTL propagator have been worked out
previously in Ref.~\cite{Peshier:1998dy}, however without discussing
renormalization scale dependences that affect any approximate
result in a quantum field theory. We also give full details
on the required analytical continuations, and we include
finite quark chemical potential.\footnote{An extension of the large-$N_{f}$ gluon self-energy to finite mass,
albeit in the weak coupling regime, can be found in Ref.~\cite{Aarts:2005vc},
where transport coefficients are calculated in the large-$N_{f}$
limit.} {\sf In particular we investigate quantitatively
the effect of Friedel oscillations in the screening
of static charges at low temperature and ultrarelativistic density.}

\section{Large-$N_f$ QED and QCD}

\begin{figure}
\begin{center}\includegraphics[%
  scale=0.8]{
  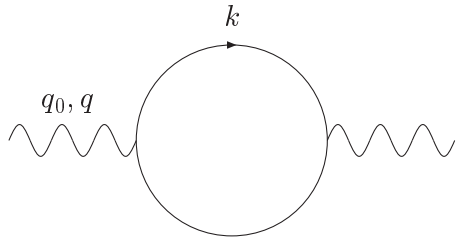}\end{center}

\caption{In the large-$N_{f}$ limit, the leading contribution ($\sim N_{f}^{0}$)
to the bosonic self-energy $\Pi_{\mu\nu}$ is given by this diagram
with one fermion loop insertion. \label{fig:bosonselfenergy}}
\end{figure}

In the large-$N_{f}$ limit, an effective coupling $\geff^{2}=g^{2}N_{f}/2$
(for QCD; $\geff^{2}=e^{2}N_{f}$ for QED) is introduced which is
kept of the order $\geff^{2}\sim O(1)$ as $N_{f}\rightarrow\infty$
and $g^{2}\rightarrow0$. All gluon-gluon interactions of QCD are
suppressed by inverse powers of $N_{f}$ -- the theory becomes QED
like. Also quark-gluon (or electron-photon) interactions are suppressed
unless they form a new fermion loop. The leading contribution to the
gauge boson self energy is therefore simply given a fermion loop
(Fig.~\ref{fig:bosonselfenergy}), which
is trivially gauge independent.
{\sf Dyson resummation of this contribution turns the gauge
boson propagator into a nonperturbative object.}

At large $N_{f}$, QCD loses asymptotic freedom. The scale dependence
of the coupling in this limit is completely determined by the one-loop
beta function\begin{equation}
\beta(\geff^{2})\equiv\frac{\bar{\mu}\partial\geff^{2}}{\partial\bar{\mu}}=\frac{\left(\geff^{2}\right)^{2}}{6\pi^{2}}\,.\label{eq:betafunction}\end{equation}
Further corrections are suppressed by at least one power of $1/N_{f}$.
Integrating this differential equation yields\begin{equation}
\frac{1}{\geff^{2}(\bar{\mu})}=\frac{1}{\geff^{2}(\bar{\mu}')}+\frac{\ln(\bar{\mu}'/\bar{\mu})}{6\pi^{2}},\label{eq:betaintegrated}\end{equation}
which implies a Landau pole at $\Lambda_{L}\sim\bar{\mu}e^{6\pi^{2}/\geff^{2}(\bar{\mu})}$.

The presence of a Landau pole means that large-$N_f$ QED and QCD
only exist as a cutoff theory. However, at finite temperature and chemical
potential, where one is primarily interested in thermal effects,
the presence of a cutoff can be neglected as long as $T,\mu \ll
\Lambda_{L}$. This requires that $\geff^2$ with renormalization scale
of the order of temperature $T$ or chemical potential $\mu$, respectively,
has to remain smaller than $\sim 36$, giving enough room for
testing the extension of hard-thermal-loop expansions from
weak to stronger coupling.\footnote{The consequences of a Landau pole
in thermal scalar field theory have been studied previously
in Refs.\ \cite{Drummond:1997cw,deVega:2000mf}.}

\section{Large-$N_f$ gauge boson propagator}

For definiteness,
we shall use the Coulomb gauge and decompose the gauge propagator
into a longitudinal and a transverse contribution\begin{eqnarray}
G_{00}(Q) & = & G_{{\rm L}}(Q)\,,\nonumber \\
G_{ij}(Q) & = & \left\{ \delta_{ij}-\frac{q_{i}q_{j}}{q^{2}}\right\} G_{T}(Q)\,,\end{eqnarray}
where the corresponding self-energy components are defined through\begin{eqnarray}
G_{L}(Q) & = & \frac{-1}{q^{2}+\Pi_{{\rm L}}(Q)}\,,\nonumber \\
G_{T}(Q) & = & \frac{1}{-Q^{2}+\Pi_{{\rm T}}(Q)}\,,\label{eq:GLGT}\end{eqnarray}
 with Minkowski metric $(+,-,-,-)$ for the 4-dimensional momenta
$Q=(q_{0},\q)$, i.e.~$Q^{2}=q_{0}^{2}-q^{2}$. The corresponding
spectral functions are determined through the discontinuity of the
Feynman propagator along the Minkowski axis \begin{equation}
\rho(q_{0},q)\equiv\frac{1}{i}\left(G(q_{0}+i\epsilon,q)-G(q_{0}-i\epsilon,q)\right).\label{eq:spectraldefinitionintro}\end{equation}
Propagating modes correspond to peaks in the spectral function. In
the case of HTL, those peaks are given by infinitely narrow delta functions,
but beyond the HTL approximation the
peaks will acquire a finite width. Such peaks usually\footnote{See
however Ref.~\cite{Blaizot:1997kw} for an interesting counterexample.}
indicate
the vicinity of a pole of the propagator to the real axis, 
but since the propagator
does not exhibit a pole in the physical sheet, we have to extend our
search to the unphysical sheet by analytic continuation. The position
of this pole can be extracted numerically from the zero of the inverse
propagator by specifying either $q$ and extracting $q_{0}(q)$ from
$G^{-1}(q_{0}(q),q)=0$ or by specifying $q_{0}$ and solving $G^{-1}(q_{0},q(q_{0}))=0$.
In the former case, we obtain information about the frequency $\omega(q)=\re\, q_{0}(q)$
and the decay constant $\gamma(q)=\im\, q_{0}(q)$ for a propagating
mode with given real-valued wave-vector $q$, which directly appears
as a peak in the corresponding spectral function. In the latter case,
we obtain the complex wave-vector $q(\omega)$ of a mode induced by
a perturbation with a given real-valued frequency $\omega$. 
In the static limit $\omega\rightarrow0$
we can extract the Debye mass from the longitudinal component $G_{L}^{-1}(q_{0}\rightarrow0,im_{D})=0$.
The transverse component is not screened in the static case, but exhibits
dynamical screening for non-zero frequency.

\subsection{Analytic continuation and scale dependence}

The main technical difficulty in finding the poles of the propagator
of Eq. (\ref{eq:GLGT}), or correspondingly the zeros of the inverse
propagator, is the correct analytic continuation of the self-energies
from the physical sheet into the neighbouring unphysical sheet. The
propagator $G(\omega+i\epsilon,q)$ neither contains poles
for $\epsilon>0$ nor for $\epsilon<0$, but these two analytic regions
are separated by a discontinuity along $\epsilon=0$. In order to
obtain the correct pole, one has to analytically continue the function
$G(\omega+i\epsilon,q)$ across this separating line from $\epsilon>0$
to $\epsilon<0$. 

The self-energy components that appear in the propagator (\ref{eq:GLGT})
can be split into a thermal piece and a vacuum piece \begin{eqnarray}
\Pi_{{\rm L}}(Q) & = & \Pi_{{\rm L,th}}(Q)-\frac{q^{2}}{Q^{2}}\Pi_{{\rm vac}}(Q)\,,\nonumber \\
\Pi_{{\rm T}}(Q) & = & \Pi_{{\rm T,th}}(Q)+\Pi_{{\rm vac}}(Q)\,,\end{eqnarray}
where the thermal pieces can be expressed as one-dimensional integrals
(see Appendix \ref{sec:Pi}), while the vacuum part depends on a renormalization
scale $\bar{\mu}$, and in the modified minimal subtraction ($\overline{{\rm MS}}$)
scheme is given by
\begin{equation}
\Pi_{{\rm \vac}}(Q)=\frac{\geff^{2}}{12\pi^{2}}Q^{2}\left\{ \log\left(\frac{-Q^{2}}{\muMS^{2}}\right)-\frac{5}{3}\right\} .\label{PivaclargeNf}\end{equation}
The analytic continuation of this vacuum piece is straightforward,
as one just has to move the branch cut of the logarithm appropriately.
One possibility is to split up the logarithm $\log(-Q^{2})\rightarrow\log(q+q_{0})+\log(q-q_{0})$.
As the logarithm has a branch cut along the negative real axis, we
would cross it for $\re\, q\pm\omega<0$ when continuing $q_{0}=\omega+i\epsilon$
from $\epsilon>0$ to $\epsilon<0$. In the first logarithm we should
therefore place the branch cut along the negative imaginary axis,
and in the second logarithm along the positive imaginary axis. It
is useful to define \begin{eqnarray}
\log^{\uparrow}(z) & := & \log(iz)-\frac{i\pi}{2},\label{logup}\\
\log^{\downarrow}(z) & := & \log(-iz)+\frac{i\pi}{2},\label{logdown}\end{eqnarray}
where $\log(z)$ is the standard logarithmic function with a branch
cut along the negative real axis. Then $\log^{\uparrow}(z)$ has the
branch cut along the positive imaginary axis, $\log^{\downarrow}(z)$
has the branch cut along the negative imaginary axis, while they both
agree to $\log(z)=\log^{\uparrow}(z)=\log^{\downarrow}(z)$ for $\re(z)>0$.
The analytically continued vacuum contribution can then be written
by replacing\begin{equation}
\log\left(\frac{q^{2}-q_{0}^{2}}{\muMS^{2}}\right)\rightarrow\log^{\downarrow}\left(q+q_{0}\right)+\log^{\uparrow}\left(q-q_{0}\right)-\log\left(\muMS^{2}\right)\end{equation}
in equation (\ref{PivaclargeNf}). 

The analytic continuation of the thermal self-energy functions turns
out to be more involved. The expressions for $\Pi_{{\rm T,th}}$ and
$\Pi_{{\rm L,th}}$, which are given in Appendix \ref{sec:Pi}, involve
a one-dimensional integral over a real variable $k$ that has to be
evaluated numerically. One might assume that it suffices to rewrite
the logarithms appearing in the integrands, similar to the vacuum
contribution $\Pi_{{\rm vac}}$, and leave the integration over the
real variable $k$ untouched. It turns out that this prescription
does not give the correct analytic continuation for the functions
$\Pi_{{\rm T,th}}(q_{0},q)$ and $\Pi_{{\rm L,th}}(q_{0},q)$, as
one can convince oneself by observing a discontinuity in the second
derivative of the self-energies at $\epsilon=0$. Indeed, we have
to be careful that the integration path for $k$ stays away from logarithmic
singularities or branch cuts as we approach and penetrate the border
of the physical sheet. In order to obtain the proper analytic continuation
we have to deform the $k$-integration path such that crossings with
singular points from the logarithms will be avoided. Another subtlety
involved is that the logarithmic arguments first have to be rewritten
in such a way that it is possible at all to deform the integration
path satisfactorily. Details of this procedure are given in Appendix
\ref{sec:analyticcontinuation}.

For later use we note that the general form of the analytically continued
inverse transverse propagator can be written as\begin{equation}
G_{T}^{-1}(Q)=-Q^{2}+\geff^{2}T^{2}f(\frac{Q}{T})+\Pi_{\vac}(Q)\label{eq:generalpropagator}\end{equation}
where the renormalization scale $\muMS$ appears explicitly in the
temperature-independent vacuum piece $\Pi_{\vac}$ and implicitly
in the coupling $\geff^{2}=\geff^{2}(\muMS)$, but the scalar function
$f(Q/T)=\Pi_{T}(\geff,T,Q)/(\geff^{2}T^{2})$ is independent of the
renormalization scale $\muMS$ as $\Pi_{T}$ is proportional to $\geff^{2}$
in the large-$N_{f}$ limit.

In the hard thermal loop (HTL) or hard dense loop (HDL) limit \cite{Braaten:1992gm,Frenkel:1992ts},
$\Pi_{{\rm L}}$ and $\Pi_{{\rm T}}$ are given by elementary functions
(listed in Appendix \ref{sec:HTL}). The HTL limit is derived by assuming
soft external lines $Q\sim gT$. In this kinematical regime, the leading
contribution is not only gauge independent (also beyond
the large-$N_f$ limit), but also formally independent of
the renormalization scale, as the vacuum contribution $\Pi_{{\rm vac}}$
is suppressed by $g^{2}$ compared to the leading HTL result. This
leading piece is of the order of the Debye mass (HTL/HDL quantities
will be marked by a hat)\begin{equation}
\hat{m}_{D}^{2}=\geff^{2}\left\{ \frac{T^{2}}{3}+\frac{\mu^{2}}{\pi^{2}}\right\} ,
\label{eq:mDlargeNfhat}\end{equation}
which is the only scale in the HTL/HDL propagator. The
HTL plasma frequency, the frequency
above which there exist propagating longitudinal and transverse modes,
is 
given by
$\omegaplhat^{2}=\hat{m}_{D}^{2}/3$; the effective mass
of transverse modes in the limit
of large momenta (asymptotic thermal mass)
by $\hat m_\infty^2=\hat{m}_{D}^{2}/2$.

In the large-$N_{f}$ limit, there is no such simple scaling, and
the shape of the dispersion relation could in principle depend on
the coupling, the temperature, and the renormalization scale. We have
a restriction through the exact renormalization dependence given by
Eq.~(\ref{eq:betafunction}) which means that we can basically choose
a scale for plotting the dispersion relations as the result does not
change if we vary renormalization scale and coupling according to
the $\beta$-function. 

To see how this works, let us assume that we know the position of
a pole of $G(Q)$, i.e.~$G^{-1}(Q)=0$. We want to show that $\mubar\partial G^{-1}(Q)/\partial\bar{\mu}=0$
is also true, provided that $\geff$ and $\bar{\mu}$ follow the $\beta$-function
$\beta(\geff^{2})\equiv\bar{\mu}\,\partial\geff^{2}/\partial\bar{\mu}$.
We find that the derivative of (\ref{eq:generalpropagator})\begin{eqnarray}
\mubar\frac{\partial G_{T}^{-1}(Q)}{\partial\mubar} & = & 0+\beta(\geff^{2})\left(T^{2}f(\frac{Q}{T})+\frac{Q^{2}}{12\pi^{2}}\left(\ln\frac{-Q^{2}}{\muMS^{2}}-\frac{5}{3}\right)\right)+\frac{\geff^{2}Q^{2}}{12\pi^{2}}(-2)\nonumber \\
 & = & \beta(\geff^{2})\left(+\frac{Q^{2}}{\geff^{2}}\right)-\frac{\geff^{2}Q^{2}}{6\pi^{2}}=0\end{eqnarray}
vanishes indeed, where we used $G_{T}^{-1}(Q)=0$ to get from the
first to the second line, and the large-$N_{f}$ $\beta$-function
Eq.~(\ref{eq:betafunction}) in the last line. Stated the other way
round, we could have derived the $\beta$-function by demanding that
$\partial G^{-1}(Q)/\partial\mubar=0$ vanishes. A similar proof holds
of course for the longitudinal component as well. Also the solution
$\omega(q)$ given implicitly by $G^{-1}(\omega(q),q)=0$ is independent
of $\muMS$. Since the Landau pole is also a solution to the zeros
of the inverse propagator $G^{-1}(\Lambda_{{\rm Landau}})=0$, the
position of the Landau pole also obeys the renormalization group equation.
The same is also true for the Debye mass, which is the solution at
zero frequency $\omega$ and purely imaginary $q$ for the longitudinal
propagator $G_{L}^{-1}(\omega=0,q=im_{D})=0$.

Therefore, if we plot the large-$N_{f}$ dispersion relations for
a specific choice of the renormalization scale, we actually plot the
curve for all couplings and renormalization scales that are connected
to each other by the renormalization group equation. The only problem
that we might run into is the presence of the Landau pole.

\subsection{Sum rules including the Landau pole}

Because of the presence of the Landau pole
in the large-$N_f$ theory, the basic spectral function sum
rule\begin{equation}
G(\omega,q)=\int_{-\infty}^{\infty}\frac{dq_{0}}{2\pi}\frac{\rho(q_{0},q)}{q_{0}-\omega},\label{eq:spectralsumrule}\end{equation}
valid for complex $\omega$, has to be modified. In the following,
we will calculate the missing contribution. Let
us first look at the zero temperature gauge field propagator. At zero
temperature and zero chemical potential, $\Pi_{T}(q_{0},q)$ and $\Pi_{L}(q_{0},q)$
vanish, and one can analytically calculate the position of the Landau
pole of the propagator (\ref{eq:GLGT}) using only the vacuum self-energy
(\ref{PivaclargeNf}). Solving for $G_{T}^{-1}(q_{0},q)=0$ we obtain\begin{equation}
q^{2}-q_{0}^{2}=\muMS^{2}\exp\left(\frac{12\pi^{2}}{\geff^{2}}+\frac{5}{3}\right)\equiv\Lambda_{L}^{2}.\label{LandauDefinition}\end{equation}
We define the spectral function as the discontinuity of the boson
propagator along the real $q_{0}$ axis\begin{eqnarray}
\rho_{T}(q_{0},q) & \equiv & \frac{1}{i}\left(G(q_{0}+i\epsilon,q)-G(q_{0}-i\epsilon,q)\right)\nonumber \\
 & = & \frac{1}{i}\left(G(q_{0}+i\epsilon,q)-G^{*}(q_{0}+i\epsilon,q)\right)\nonumber \\
 & = & 2\im G(q_{0}+i\epsilon,q)\,.\label{eq:spectraldefinition}\end{eqnarray}
The integral over the spectral function\begin{equation}
\int_{-\infty}^{\infty}\frac{dq_{0}}{2\pi}\frac{\rho(q_{0},q)}{q_{0}-\omega}=\int_{-\infty}^{\infty}\frac{dq_{0}}{2\pi}\left(\frac{G(q_{0}+i\epsilon,q)}{q_{0}-\omega}-\frac{G(q_{0}-i\epsilon,q)}{q_{0}-\omega}\right)\end{equation}
 can be calculated by closing the contour by a harmless grand half
circle at infinity and picking up all pole contributions. 
The
Landau pole (\ref{LandauDefinition}) provides
additional poles in the $q_0$ plane
\begin{equation}
q_{0}=\pm i\sqrt{\Lambda_{L}^{2}-q^{2}}\equiv\pm i\Lambda_{L}(q).\end{equation}

\begin{figure}
\begin{center}\includegraphics[%
  scale=0.7]{
  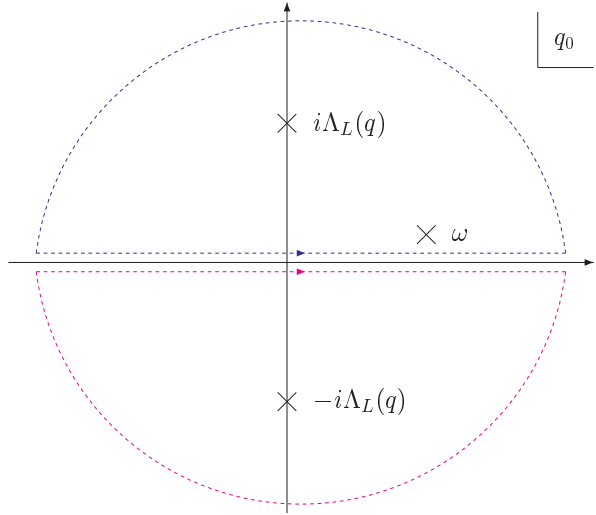}\end{center}

\caption{Integration path for the spectral function with three poles: One
of those originates from the denominator $1/(q_{0}-\omega)$, while
the other two poles are the unavoidable Landau poles.\label{cap:integrationpath}}
\end{figure}
The
first pole lies in the upper half of the complex plane and 
its residue can be
calculated as:\begin{equation}
{\rm Res}_{q_{0}\rightarrow i\Lambda_{L}(q)}\frac{G(q_{0}+i\epsilon,q)}{q_{0}-\omega}=\frac{6\pi^{2}}{\geff^{2}}\frac{1}{i\Lambda_{L}(q)-\omega}\frac{1}{i\Lambda_{L}(q)}.\end{equation}
The pole in the lower half $q_{0}=-i\Lambda_{L}(q)$ will receive
an additional minus sign from the integration orientation, so that
the total contribution (including the usual pole) is given by\begin{equation}
\int_{-\infty}^{\infty}\frac{dq_{0}}{2\pi}\frac{\rho(q_{0},q)}{q_{0}-\omega}=G(\omega+i\epsilon,q)-\frac{12\pi^{2}}{\geff^{2}}\frac{1}{\Lambda_{L}^{2}-q^{2}+\omega^{2}}\label{eq:spectralcorrectionvacuum}\end{equation}
where we have substituted back $\Lambda_{L}^{2}(q)=\Lambda_{L}^{2}-q^{2}$
with $\Lambda_{L}$ as defined in (\ref{LandauDefinition}). 


With a full one-loop self energy insertion,
the position of the Landau pole $\Lambda_{L}(q)$ as the solution
to $G^{-1}(q_{0}=\Lambda_{L}(q),q)=0$ for a given $q$ can in general
only be calculated numerically. In the previous section we simply
had $\Lambda_{L}(q)=\sqrt{\Lambda_{L}^{2}-q^{2}}$, but the dependence
on $q$ is more complicated in the general case. Following the previous
derivation for a general propagator, we can write the residue at the
Landau pole as\begin{equation}
{\rm Res}_{q_{0}\rightarrow i\Lambda_{L}(q)}\frac{G(q_{0}+i\epsilon,q)}{q_{0}-\omega}=\frac{1}{i\Lambda_{L}(q)-\omega}\left(\left.\frac{\partial G^{-1}(q_{0},q)}{\partial q_{0}}\right|_{q_{0}=i\Lambda_{L}(q)}\right)^{-1}.\end{equation}
Using the pole in the lower half plane and the property that the derivative
of $G(q_{0},q)$ is antisymmetric along the Euclidean axis  $\left.\partial G^{-1}(q_{0},q)/\partial q_{0}\right|_{q_{0}=-i\Lambda_{L}(q)}$$=-\left.\partial G^{-1}(q_{0},q)/\partial q_{0}\right|_{q_{0}=i\Lambda_{L}(q)}$
($G(q_{0},q)$ is real and symmetric on the Euclidean axis, so its
derivative is antisymmetric), the full relation is then given by\begin{equation}
\int_{-\infty}^{\infty}\frac{dq_{0}}{2\pi}\frac{\rho(q_{0},q)}{q_{0}-\omega}=G(\omega+i\epsilon,q)-\frac{2\Lambda_{L}(q)}{\Lambda_{L}(q)^{2}+\omega^{2}}\left(\left.\frac{\partial G^{-1}(q_{0},q)}{i\partial q_{0}}\right|_{q_{0}=i\Lambda_{L}(q)}\right)^{-1}\label{eq:spectralsumrulecorrected}\end{equation}
where the last part is real, despite the appearance of explicit {}``$i$''s.
The last term corrects the usual spectral function sum rule due to
the presence of the Landau pole. A rough estimate of the size of this
correction can be obtained from the last term of (\ref{eq:spectralcorrectionvacuum}):
For small $Q^{2}$ it is around $5\times10^{-7}T^{-2}$ for $\geff^{2}(\muMS=\pi T)=9$
and around $2\times10^{-3}T^{-2}$ at $\geff^{2}(\pi T)=36$, and
increases with increasing $Q^{2}$. For soft momenta $Q\sim\geff T$
one could estimate roughly that this term gives a correction to the
propagator of the order of $(T/\Lambda_{L})^{2}$. For hard momenta
$Q\sim T$ the correction is of order $(T/\geff\Lambda_{L})^{2}$.

\section{Results}

\subsection{Dispersion relations at zero chemical potential}

\begin{figure}
\begin{center}\includegraphics{
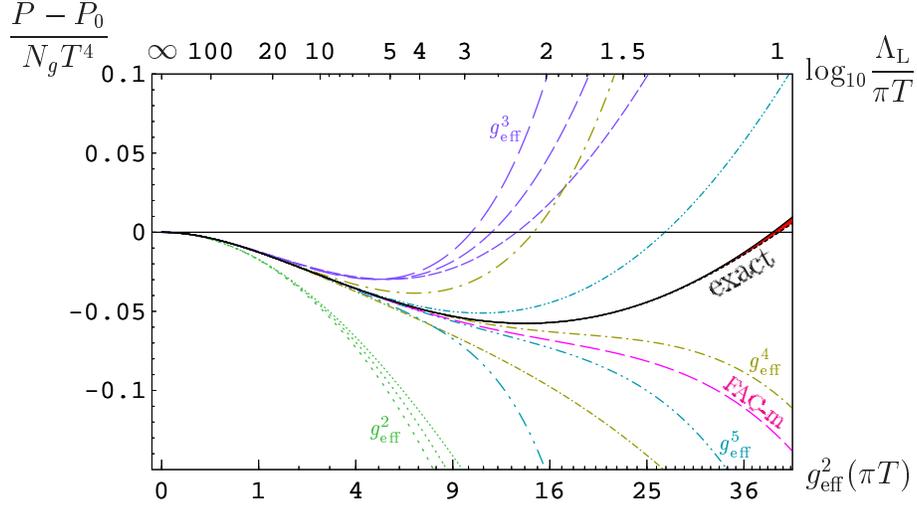}\end{center}

\caption{Exact result for the interaction pressure at finite temperature and
zero chemical potential at large $N_{f}$ as a function of $\geff^{2}(\muMS=\pi T)$,
compared to strict perturbation theory \cite{Moore:2002md,Ipp:2003zr,Ipp:2003jy}.
The tiny band appearing for large values of the coupling for the exact
result shows the cutoff dependence from varying the upper numerical
integration cutoff between $1/\sqrt{4}=$50\% and $1/\sqrt{2}\approx$70\%
of the Landau pole $\Lambda_{L}$. The results of strict perturbation
theory are given through order $\geff^{2}$ (dotted line), $\geff^{3}$
(dashed), $\geff^{4}$ (dash-dotted), and $\geff^{5}$ (dash-dot-dotted)
where the renormalization scale $\muMS$ is varied between $\frac{1}{2}\pi T$
(line pattern slightly compressed), $\pi T$, and $2\pi T$ (line
pattern slightly stretched). The line labelled {}``FAC-m'' indicates
the scale chosen by the prescription of fastest apparent convergence
as indicated in the text for which the curves of $\geff^{4}$ and
$\geff^{5}$ coincide. \label{fig:pressure}}
\end{figure}

Figure \ref{fig:pressure} recalls
the pressure obtained in the large-$N_{f}$ limit 
for finite temperature and
zero quark chemical potential \cite{Moore:2002md,Ipp:2003zr,Ipp:2003jy}.
The full line gives the result at large
$N_{f}$. The ambiguity introduced by varying the numerical cutoff
below the Landau pole $Q_{{\rm max}}^{2}=a\Lambda_{L}^{2}$ between
$a=1/4$ and $1/2$ is shown as a tiny (red) band for large couplings.
This ambiguity is suppressed by a factor $(T/\Lambda_{L})^{4}$. 
Practically, this means that we can study a range of couplings 
$\geff^{2}(\muMS=\pi T)\lesssim36$
safely, where this number of course depends on the choice
of renormalization scale $\muMS$ [a change
of this scale changes the coupling according to 
Eq.~(\ref{eq:betaintegrated})]. Perturbation
theory ceases to work much sooner and shows large scale dependences
for couplings already at $\geff^{2}\gtrsim4$. Using an optimized
renormalization scale, namely fastest apparent convergence of the effective
mass parameter $m_{{\rm E}}$ (FAC-m), the perturbative result
to order $\geff^{5}$ follows the exact result up to $\geff^{2}\approx8$.
Still, 
there is a large range of couplings $8\lesssim\geff^{2}(\pi T)\lesssim36$
to explore where strict perturbation theory fails, but the ambiguity introduced
by the vicinity of the Landau pole is still negligible.

One of the remarkable and so far 
not easily understood features of
the pressure curve is the minimum appearing at a coupling $\geff^{2}(\pi T)\approx14$.
Paradoxically, in this strong coupling range $\geff^{2}\gtrsim14$
the pressure seems to approach again the interaction-free pressure
value. 
{\sf Since the nontrivial content of the
large-$N_f$ pressure can be expressed entirely in terms
of the gauge boson propagator \cite{Moore:2002md,Ipp:2003zr,Ipp:2003jy},
it is natural to ask whether
this nonmonotonic behavior is associated with 
qualitative changes in the properties of the gauge boson propagator.
We shall find that this is not really the case. As will be shown
in a subsequent paper \cite{Birr}, the nonmonotonic behavior is
instead associated with nonmonotonic behavior of the asymptotic
quark mass, which is a required ingredient when expressing the
entropy in terms of quasiparticle spectral
data \cite{Vanderheyden:1998ph,Blaizot:2000fc}.
}

In the following we will plot spectral functions and dispersion relations
covering weak and strong coupling for the gauge boson propagator in
the large-$N_{f}$ limit, and compare them to the corresponding HTL
results. For larger couplings, the question of renormalization scale
dependence becomes important. In principle one could plot the following
comparisons between exact large-$N_{f}$ and HTL results at various
scales. To avoid cluttering the plots, we decided to scale large-$N_{f}$
quantities by the large-$N_{f}$ Debye mass $m_{D}$, and HTL quantities
by the HTL Debye mass $\hat{m}_{D}$ and present them in the same
plot. Each of the quantities is then separately renormalization scale
independent, while the renormalization
scale dependence of the corresponding HTL approximations
can be read off separately from a plot $m_{D}/\hat{m}_{D}$.
(Note that the plasma frequency would be less appropriate to set a
scale of the order $\sim gT$ because of the imaginary part it obtains
at larger couplings $\geff^{2}$ -- the Debye mass on the contrary
always stays purely real.) For the spectral functions we note that
$\geff^{2}G(q_{0},q)$ and therefore $\geff^{2}\rho(q_{0},q)$ is
a renormalization scale independent quantity. 


%
\begin{figure}
\begin{center}\includegraphics{
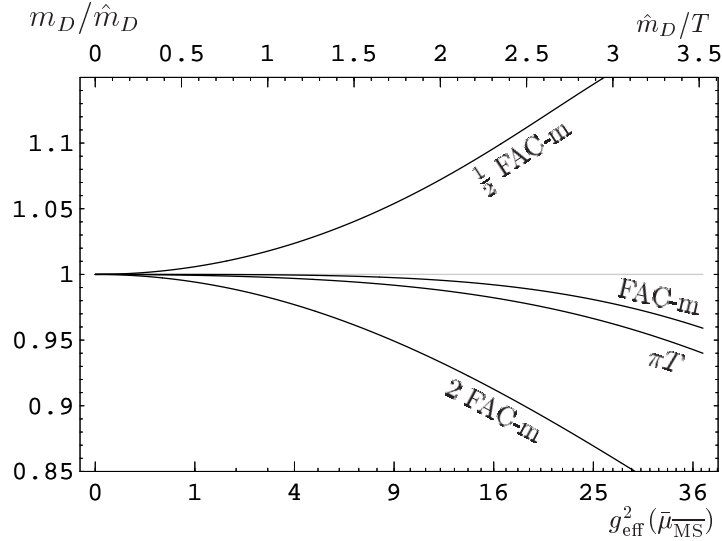}\end{center}

\caption{Comparison of the large-$N_{f}$ Debye mass $m_{D}$ to the corresponding
HTL value $\hat{m}_{D}$. The renormalization scale $\muMS$ is varied
by a factor of 4 around FAC-m. Using FAC-m or $\pi T$ as the renormalization
scale brings the HTL result within a few percent of the full result
for the whole range of couplings $\geff^{2}$ plotted. Note that the
abscissa does not denote the coupling at a fixed scale, but at a different
scales as indicated on the lines. \label{fig:debyemass}}
\end{figure}

One of the most prominent properties of the gauge boson propagator
is the emergence of a screening mass in the static longitudinal
sector, the Debye mass. 
Numerically, we obtain the Debye mass as the solution to 
Eq.~(\ref{eq:GLGT})
in the static limit,
\begin{equation}
q^{2}+\Pi_{{\rm L}}(q_{0}=0,q)=0\qquad{\rm at\,}q=im_{D}.\end{equation}
While in the HTL approximation $\hat\Pi_{{\rm L}}(q_{0}=0,q)=\hat m_D^2$
is a constant given by Eq.~(\ref{eq:mDlargeNfhat}), this is no longer the
case for the complete expression
in the large-$N_{f}$ limit. Nevertheless,
the solution for $m_D$ indeed turns out to be a purely real number.
The latter can be identified with the screening
mass in a Yukawa potential
for static sources. 
Figure~\ref{fig:debyemass} shows a comparison of the Debye
mass in the large-$N_{f}$ limit, normalized to its corresponding
HTL value (\ref{eq:mDlargeNfhat}). Other than the following plots, the
axis of Fig.~\ref{fig:debyemass} is not labelled at a fixed renormalization
scale $\geff^{2}(\pi T)$, but rather at a different scale for each
curve as denoted next to each line. The four curves are in fact related
by the 
renormalization group equation (\ref{eq:betaintegrated}),
because the position
of the large-$N_{f}$ Debye mass is renormalization scale independent,
like any singularity of the propagator. The HTL result on the other
hand is a truncated result at order $\geff^{2}$. Therefore the HTL
Debye mass does not run properly with the renormalization scale,
leading to a renormalization-scale dependence of $m_D/\hat m_D$.

For small $\geff^{2}$ an optimal choice of renormalization scale is
obtained by requiring that the $\geff^{4}$
contribution to the Debye mass $m_{D}$ for the full large-$N_{f}$
result vanishes. This prescription of fastest apparent convergence
will be labelled as FAC-m and leads to 
\cite{Moore:2002md,Ipp:2003zr,Ipp:2003jy}
\begin{equation}
\muMS^{\textrm{FAC-m}}=e^{\frac{1}{2}-\gamma_{E}}\pi T\approx0.9257\pi T,
\end{equation}
which is very close to the choice $\muMS=\pi T$,
the lowest Matsubara frequency. 
Even for the largest couplings
$\geff^{2}\approx36$ the deviation from the HTL Debye mass is only of
the order of 5\% for this choice. It will therefore suffice to compare
all following quantities at a fixed scale $\muMS=\pi T$, keeping
in mind that a change of the scale e.g.\ by a factor of 2 will introduce
additional deviations as displayed in Fig.~\ref{fig:debyemass}.

\begin{figure}
\begin{center}\includegraphics[%
  scale=0.9]{
  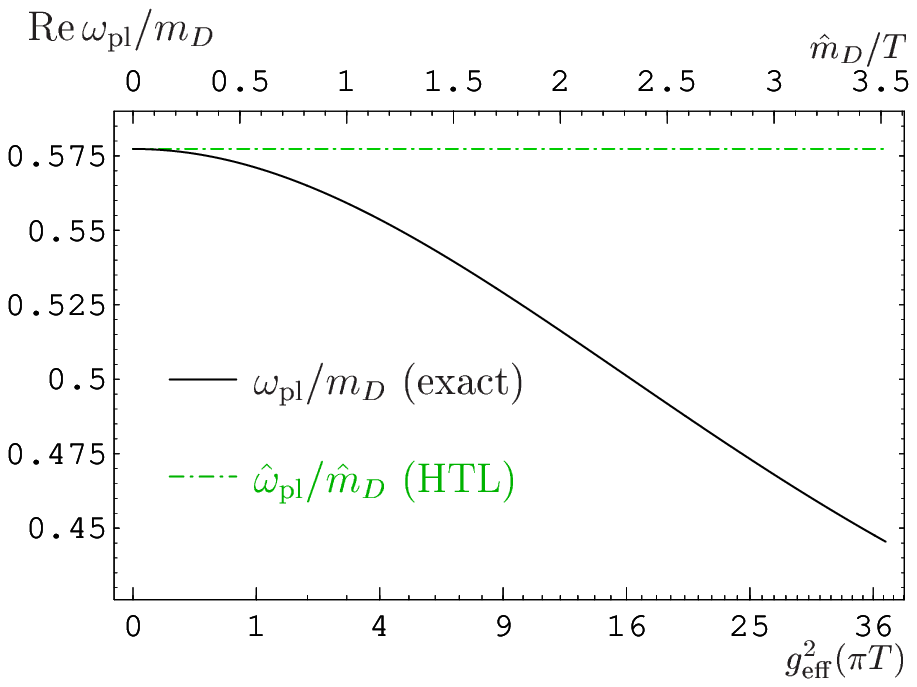}\includegraphics[%
  scale=0.9]{
  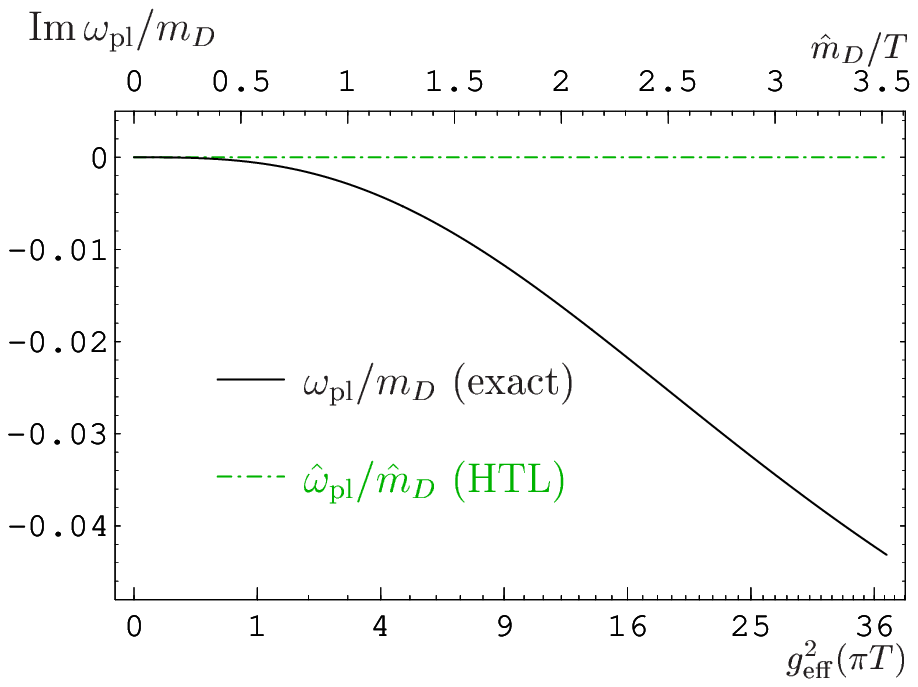}\end{center}

\caption{Real and imaginary part of the plasma frequency $\omega_{pl}$ in
the large-$N_{f}$ limit compared to the HTL value $\hat{\omega}_{pl}/\hat{m}_{D}=1/\sqrt{3}\approx0.577$
as a function of $\geff^{2}(\muMS=\pi T)$ at the renormalization
scale $\pi T$. The imaginary part of the plasma frequency, the damping
constant, is caused by the vacuum process of quark-antiquark pair
creation, and thus missing in the purely thermal HTL approximation.
\label{fig:plasmafrequency}}
\end{figure}

Another quantity of central importance is the plasma frequency: It
indicates the lowest frequency for which there exist propagating modes
in the plasma. 
Transverse and longitudinal modes share the same plasma frequency
in an isotropic plasma.
Frequencies below the plasma frequency are associated with
spatial screening.
{\sf 
The plasma frequency and the frequency of propagating modes
are obtained from
\begin{equation}
G_{L,T}^{-1}(\omega(q),q)=0,\qquad q\in {\mathbb{R}}.
\end{equation}
While the dispersion relations obtained in the HTL approximation
involve only real frequencies $\omega(q)$,
the full large-$N_f$ result is complex.
Real and imaginary parts are plotted separately in 
Fig.~\ref{fig:plasmafrequency}.
Compared to the real part, the magnitude of the imaginary part
remains small for all values of the coupling.}
For small couplings, the plasma frequency approaches
the HTL value, which is given by $\omegaplhat=\hat{m}_{D}/\sqrt{3}\approx0.57735\,\hat{m}_{D}$.
As the coupling increases, the renormalization-scale independent ratio
of plasma frequency to Debye mass decreases by 10\% at $\geff^{2}(\muMS=\pi T)=14$
and by 20\% at $\geff^{2}=36$. 
The
imaginary part is due to the vacuum process of quark-antiquark pair creation,
which is missing in the HTL approximation.
{\sf (In QCD at finite $N_f$, the damping rate of propagating
bosonic modes is dominated by Bose-enhanced gluonic contributions
proportional to a single power of $g$ rather than to
$\geff^2=g^2 N_F/2$ \cite{Braaten:1990it,Pisarski:1993rf,Flechsig:1995sk}.)}

\begin{figure}
\begin{center}\includegraphics[%
  scale=0.9]{
  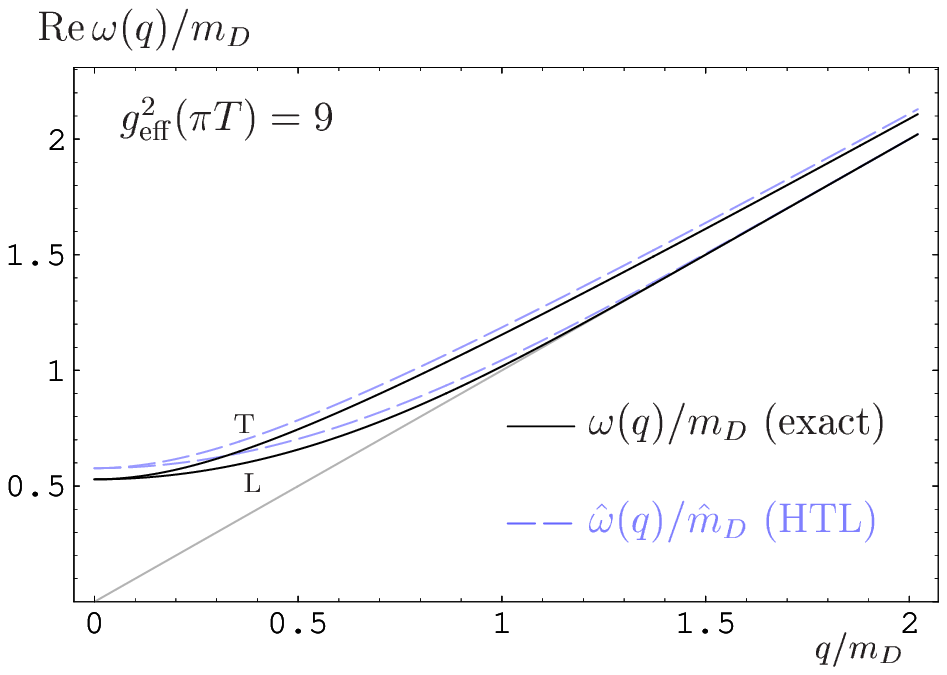}\includegraphics[%
  scale=0.9]{
  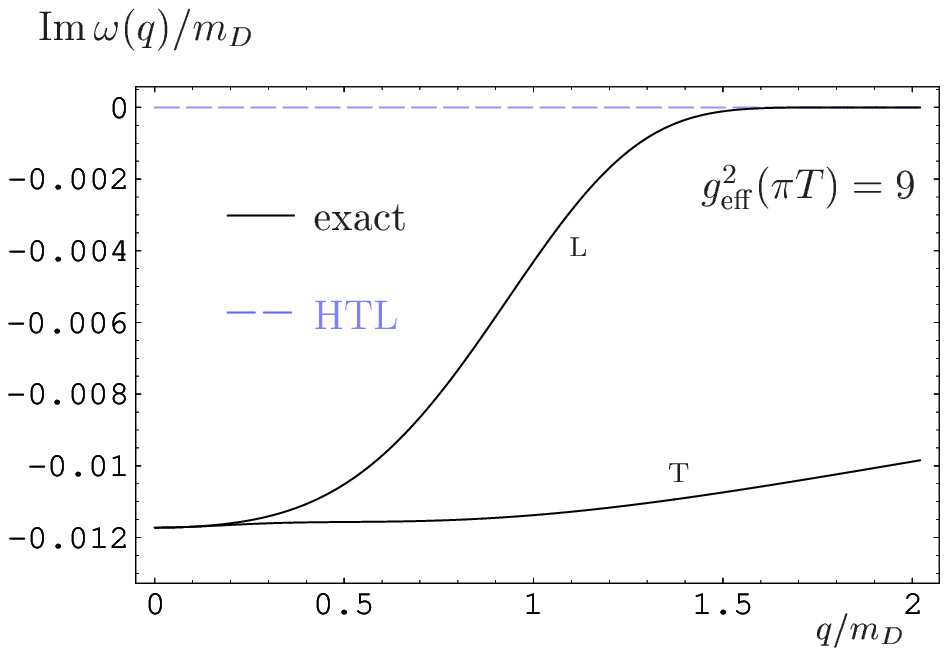}\end{center}

\caption{Real and imaginary part of the dispersion relation $\omega(q)$ as
a function of real $q$ for $\geff^{2}=9$ in the large-$N_{f}$ limit
compared to the HTL result. Abscissa and ordinate for the exact large
$N_{f}$ result are scaled by the large-$N_{f}$ Debye mass $m_{D}$,
while the HTL curve is on both axis scaled by the HTL mass $\hat{m}_{D}$.
At this moderate value of the coupling, which lies at the border of
applicability of strict perturbation theory, the plasma frequency
is slightly lowered and obtains a small imaginary part. The light
gray line indicates the lightcone. \label{fig:dispersionw}}
\end{figure}

\begin{figure}
\begin{center}\includegraphics[%
  scale=0.9]{
  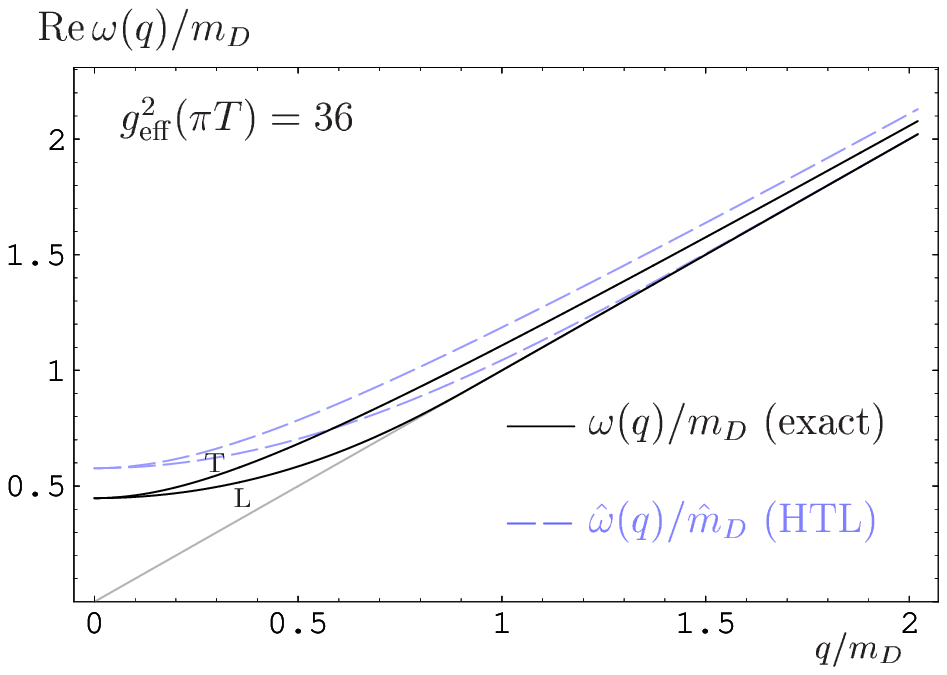}\includegraphics[%
  scale=0.9]{
  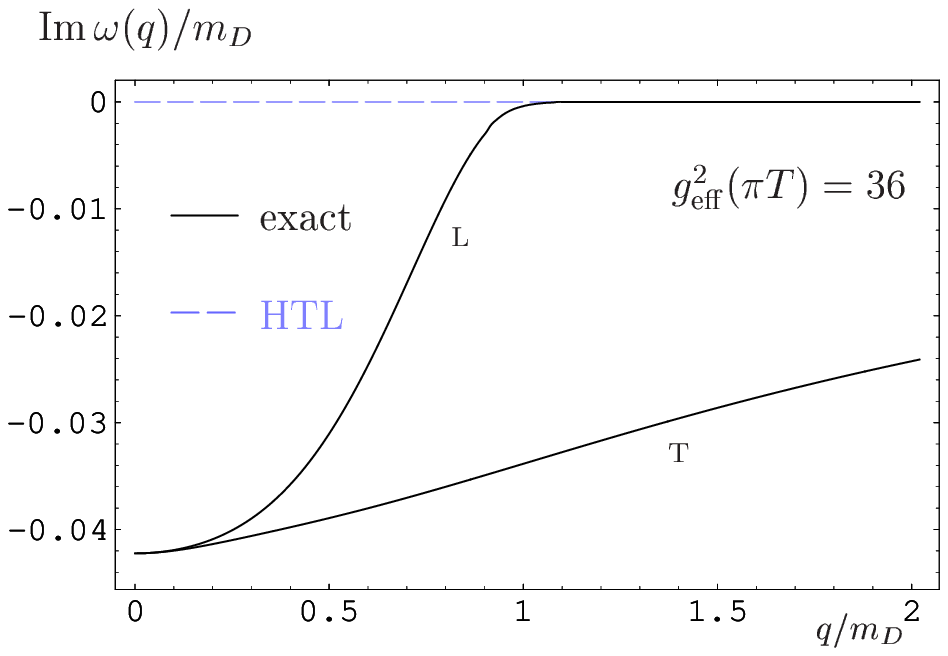}\end{center}

\caption{Same as Figure \ref{fig:dispersionw} for the larger coupling $\geff^{2}(\pi T)=36$.
Real and imaginary part are further lowered compared to the case of
$\geff^{2}(\pi T)=9$. As the longitudinal branch approaches the lightcone
faster, also its imaginary part approaches 0 at smaller values of
$q/m_{D}$. \label{fig:dispersionw36}}
\end{figure}

Figures \ref{fig:dispersionw} and \ref{fig:dispersionw36} show the
dispersion curve for representative values of the coupling $\geff^{2}(\mubar=\pi T)=9$
and 36. These plots are scaled by the large-$N_{f}$ plasma frequency
and thus show a renormalization scale independent result. HTL quantities
are scaled by the HTL Debye mass. One notes the lower plasma frequency
compared to the corresponding HTL value, albeit the effect
is moderate even at large coupling. 
The imaginary part of the
longitudinal branch vanishes quickly as the curve approaches the light-cone.
In the limit of vanishing $q/m_{D}$, 
the real as well as the imaginary parts of transverse
and longitudinal contributions coincide.\footnote{This
is due to the isotropy of the plasma. See Ref.\ \cite{Romatschke:2003ms}
for counterexamples in anisotropic plasmas within the hard-loop
approximation.} 
The unique plasma frequency
at $q=0$ evolves according to Fig.~\ref{fig:plasmafrequency}.

\begin{figure}
\begin{center}\includegraphics[%
  scale=0.9]{
  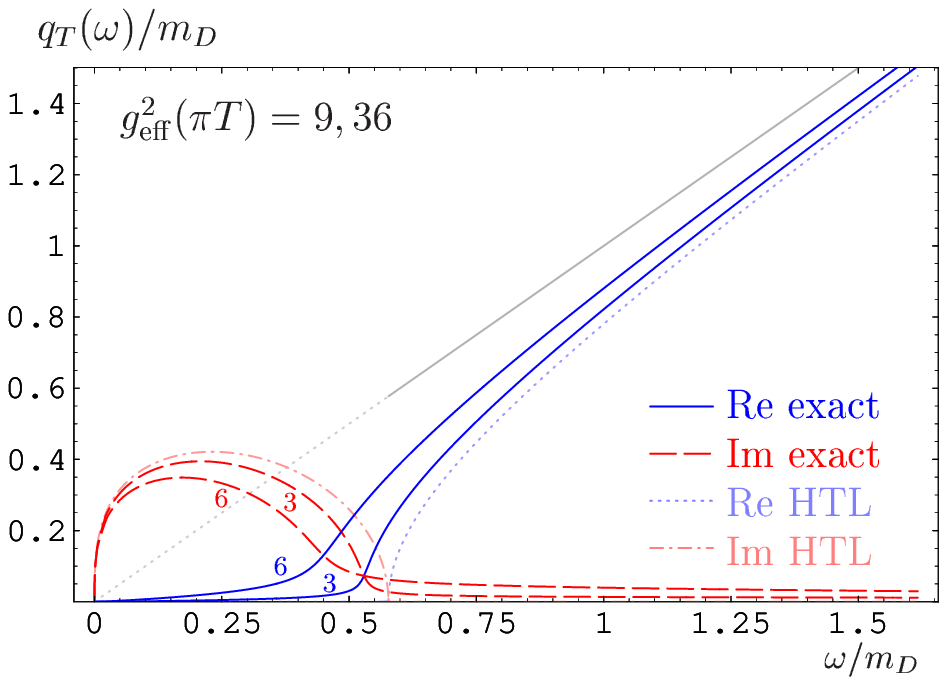}\includegraphics[%
  scale=0.9]{
  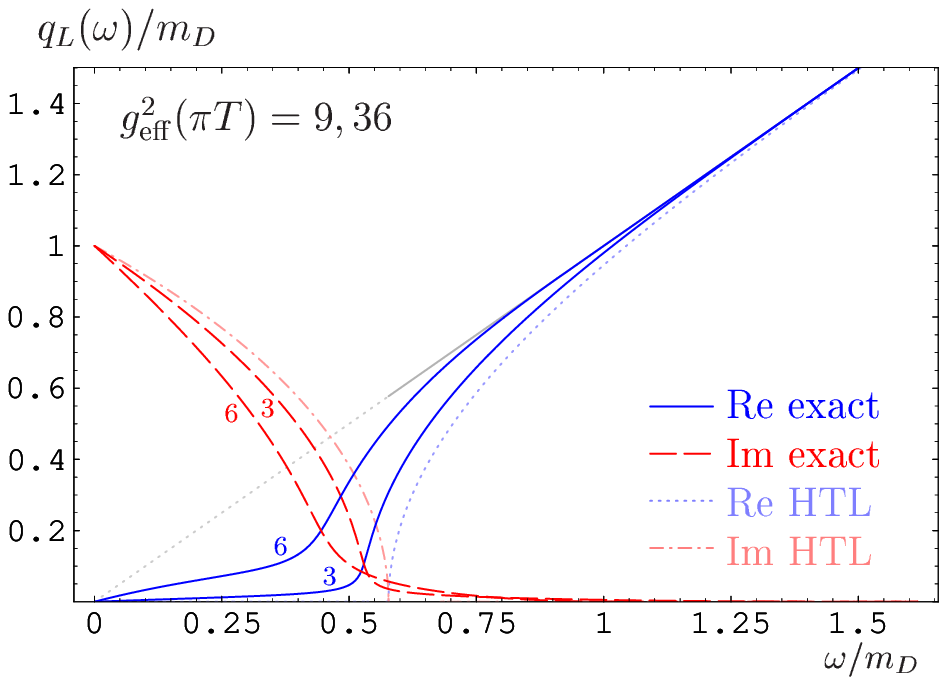}\end{center}

\caption{Real and imaginary part of the dispersion relation $q(\omega)$ as
a function of real $\omega$ for $\geff^{2}=9$ and $36$ (labeled
in the plot as 3 and 6 respectively for $\geff$) in the large-$N_{f}$
limit compared to the HTL result. The left panel shows the dispersion
relation for the transverse component $q_{T}(\omega)$, while the
right panel shows the longitudinal contribution $q_{L}(\omega)$.
The exact large-$N_{f}$ result is scaled by the large-$N_{f}$ Debye
mass $m_{D}$ while the HTL result is scaled by the HTL Debye mass
$\hat{m}_{D}$. Notably, the real part of $q(\omega)$ vanishes in
the static case $\omega\rightarrow0$ in the large-$N_{f}$ limit
also for larger couplings $\geff^{2}$, and one can see from the right
panel that the Debye mass stays a real quantity even beyond HTL. \label{fig:dispersionq}}
\end{figure}

Instead of assuming real wave vectors $q$ and solving numerically
for complex $\omega(q)$ as we had done in Fig.~\ref{fig:dispersionw},
we could start from a given real frequency $\omega$ and calculate
the corresponding complex wave vector $q(\omega)$. 
{\sf This corresponds to calculating the linear response to
perturbations with a given real frequency.}
The result is
shown in Figure \ref{fig:dispersionq} for two different couplings
$\geff^{2}(\muMS)=9$ and 36 (which are labelled as $\geff=3$ and
6 respectively in the plot). 
Above the plasma frequency, one has weakly damped spatial
oscillations, and below the plasma frequency there is predominantly
exponential screening, modulated by comparatively long-range
oscillations that are absent in the HTL case (though
fundamentally different from the
Friedel oscillations \cite{FetW:Q,Kapusta:1988fi}
we shall discuss further below).
Following the curves
down to vanishing frequency $\omega\rightarrow0$
leads to the (real) value
of the Debye mass in the longitudinal
mode (right panel of Fig.~\ref{fig:dispersionq}), 
and in the transverse mode (left panel) to unscreened
static modes (but significant
dynamical screening at small but non-vanishing $\omega$).
The real part of $q(\omega)$, which corresponds to spatial
oscillations, vanishes for both transverse and longitudinal
modes, in the static limit $\omega\rightarrow0$. The imaginary part
of the longitudinal modes for $\geff=3$, 6, and HTL, approaches the
value 1 in the static case as the curves are normalized to this value.
The HTL results of Fig.~\ref{fig:dispersionq}, which are given by
dotted and dash-dotted lines, are of course independent of the coupling
in units of the HTL Debye mass. In a proper comparison between large
$N_{f}$ and HTL one should keep the renormalization scale dependence
of $m_{D}/\hat{m}_{D}$ from Fig.~\ref{fig:debyemass} in mind.

\begin{figure}
\begin{center}\includegraphics[%
  scale=0.9]{
  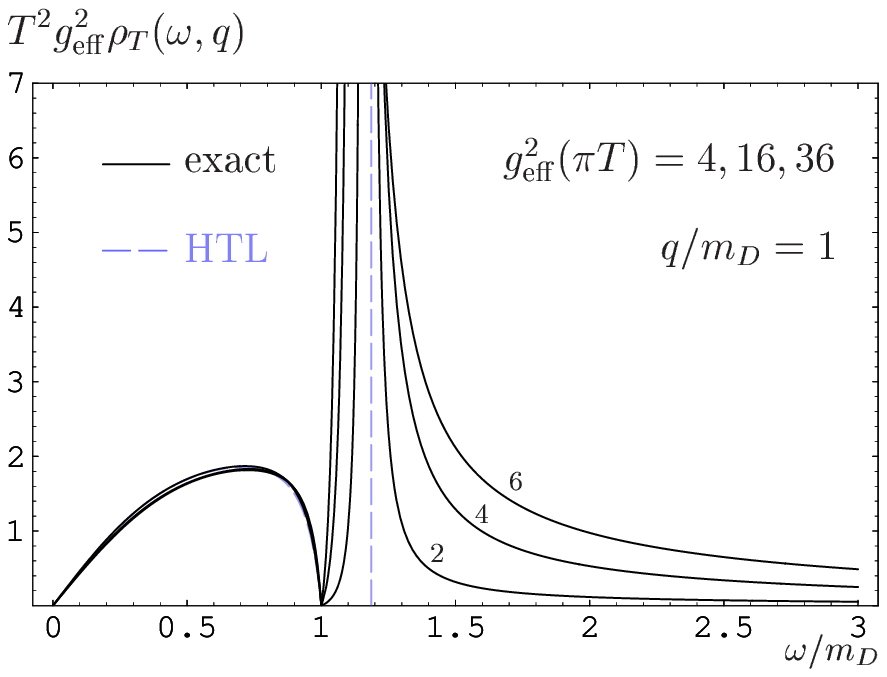}\includegraphics[%
  scale=0.9]{
  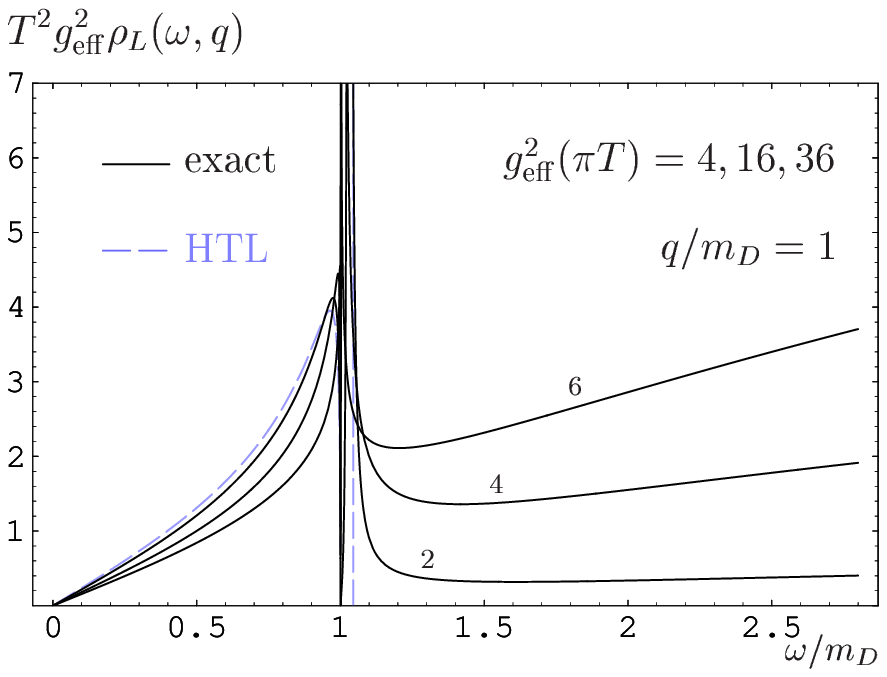}\end{center}

\caption{Spectral function $\rho_{T}$ and $\rho_{L}$ as a function of $\omega/m_{D}$
for $q/m_{D}=1$ for various couplings $\geff^{2}(\muMS=\pi T)$ (the
lines in the plot are labelled by the values for $\geff=2,4,6$).
As mentioned in the text, $\geff^{2}\rho(\omega,q)$ is a renormalization
scale independent quantity, and $T^{2}\geff^{2}$ scales like $\hat{m}_{D}^{2}$
such that the HTL results for various couplings lie on top of each
other. The position of the infinitely narrow HTL $\delta$-peak is
indicated by the vertical line, while the peaks at large $N_{f}$
have a finite width and are shifted to lower energies with increasing
coupling. \label{fig:spectral1}}
\end{figure}

\begin{figure}
\begin{center}\includegraphics[%
  scale=0.9]{
  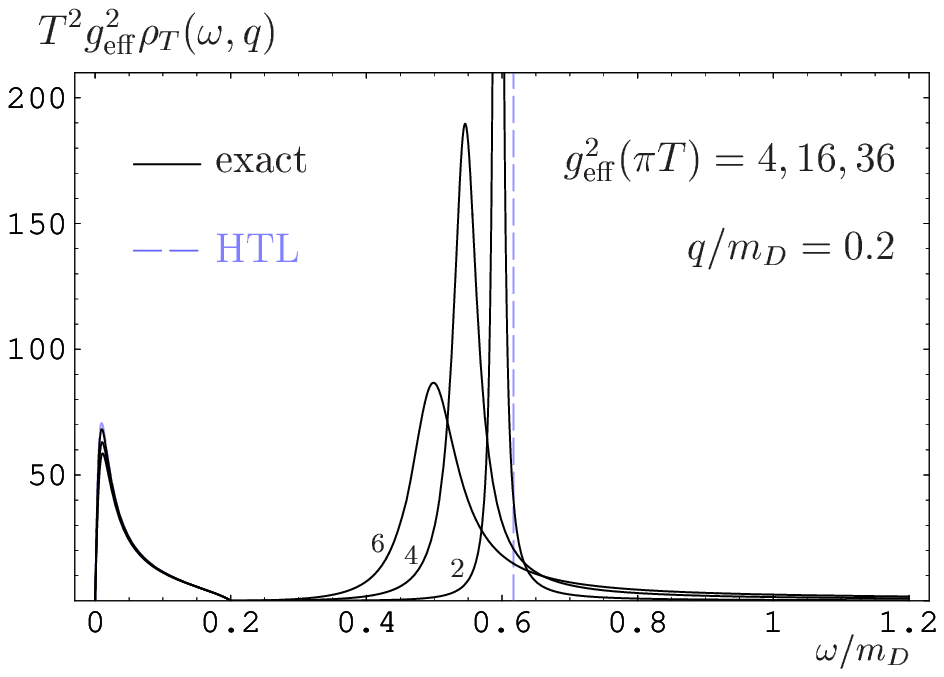}\includegraphics[%
  scale=0.9]{
  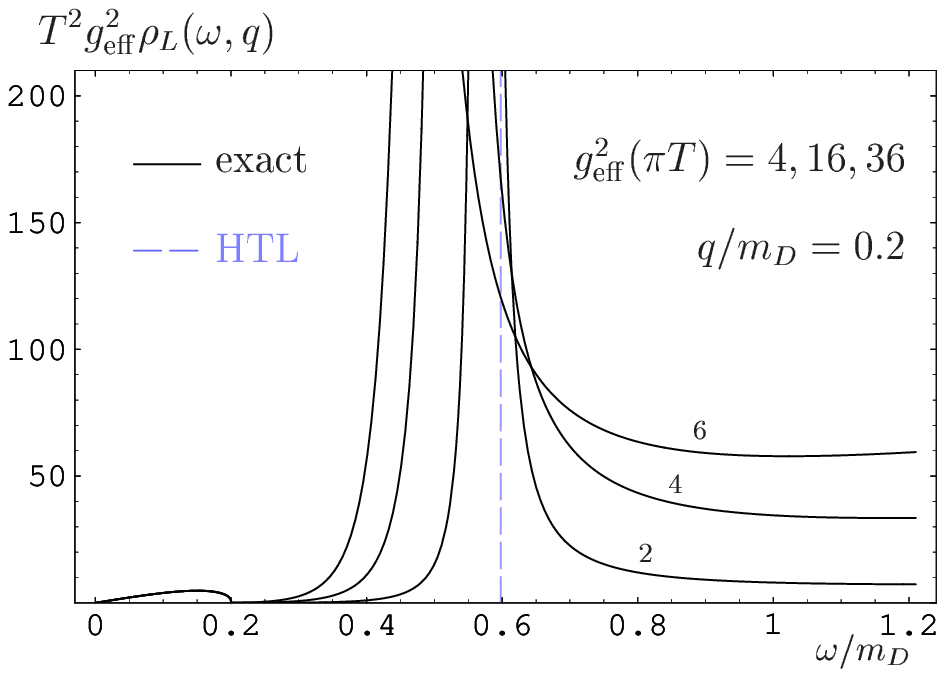}\end{center}

\caption{Same as figure (\ref{fig:spectral1}) for for $q/m_{D}=0.2$. Due
to the imaginary part of the plasma frequency, the peak of the propagating
modes broadens. Additionally their frequency lowers. \label{fig:spectral02}}
\end{figure}

Figures \ref{fig:spectral1} and \ref{fig:spectral02} show the spectral
functions at two different values of $q/m_{D}$. The combination 
$T^{2}\geff^2\rho(\omega,q)$
on the ordinate is chosen in order to show a renormalization scale
independent result: The quantity $\geff^{2}G$ is renormalization
scale independent, which can be seen by combining Equations (\ref{eq:betaintegrated})
and (\ref{eq:generalpropagator}). In all four plots the broadening
of the propagating mode peak with increasing coupling is evident.
For the smaller wave vector $q/m_{D}=0.2$ in Fig.~\ref{fig:spectral02},
one furthermore nicely observes the decrease of the plasma frequency
with increasing coupling. The HTL pole corresponds to a delta function peak, 
indicated
by the vertical line.

\subsection{Dispersion relations at finite chemical potential}

\begin{figure}
\begin{center}\includegraphics{
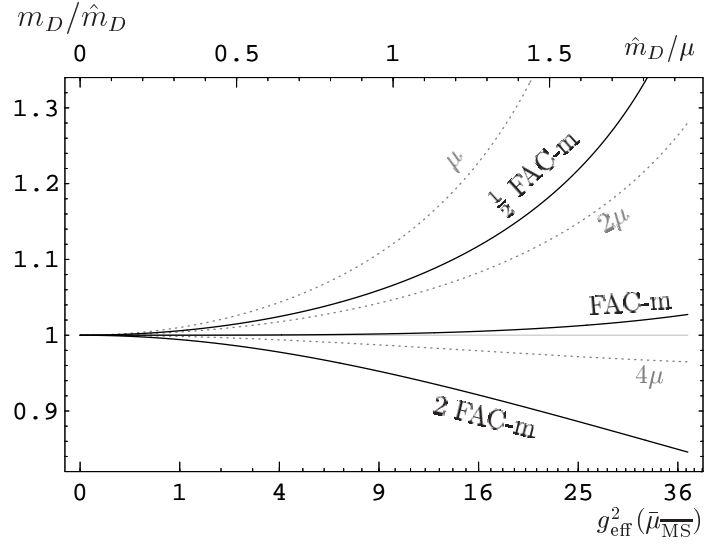}\end{center}

\caption{Comparison of the large-$N_{f}$ Debye mass $m_{D}$ to the corresponding
HDL value $\hat{m}_{D}$ at finite chemical potential and zero temperature.
The FAC-m scale takes the value $\muMS=2\sqrt{e}\mu$, and is varied
by a factor of 4 in the plot. For comparison, also the scales $\muMS=\mu$,
$2\mu$, and $4\mu$ are displayed. The full Debye mass is in remarkable
agreement with the HDL result for the FAC-m scale.\label{fig:debyemasszeroT}}
\end{figure}

In the following,
we extend the analysis to finite chemical potential $\mu$ and zero
temperature. Figure \ref{fig:debyemasszeroT} shows the corresponding
Debye mass. Here the prescription of fastest apparent convergence
gives \[
\textrm{FAC-m}=2\sqrt{e}\mu\approx3.2974\mu\qquad{\rm at\,}T\rightarrow0.\]
The FAC-m scale for arbitrary finite $T$ and $\mu$ can be found
in Ref.~\cite{Ipp:2003yz}. In addition to multiples of the FAC-m
scale, also the simple scale choice $\muMS=2\mu$, varied within a
factor of 4, is depicted.

\begin{figure}
\begin{center}\includegraphics[%
  scale=0.9]{
  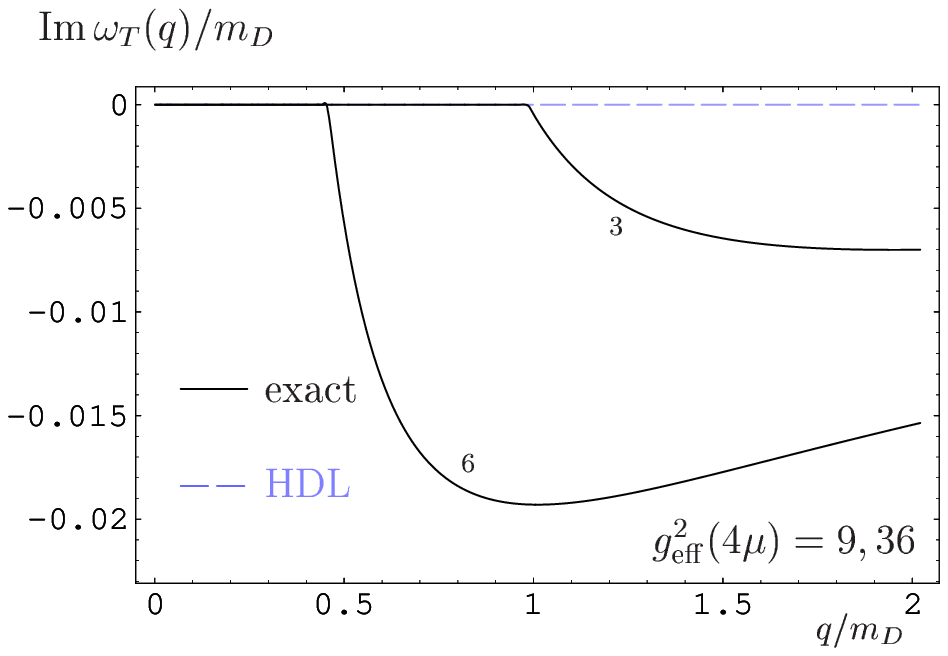}\includegraphics[%
  scale=0.9]{
  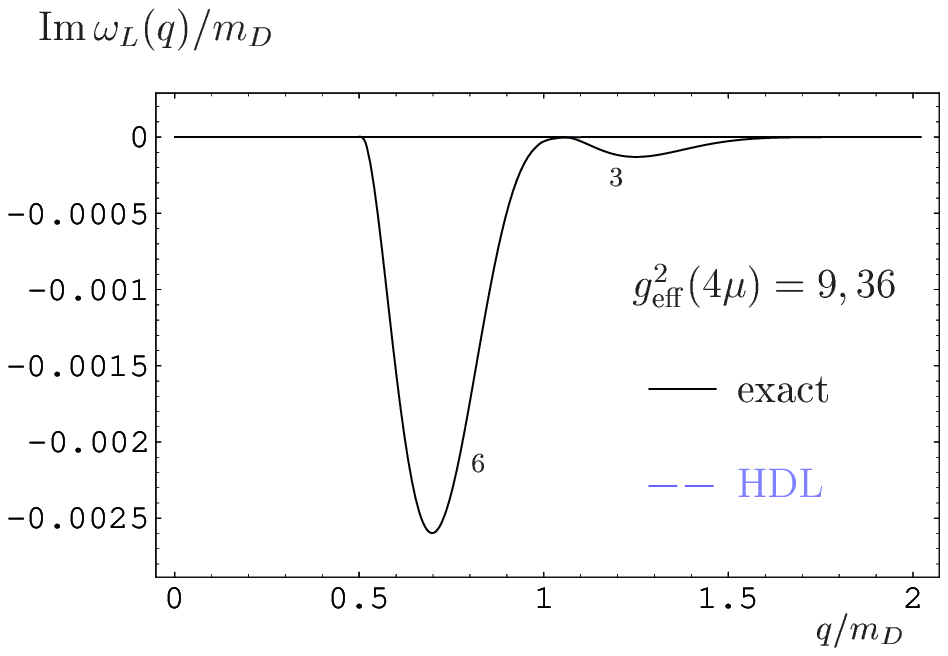}\end{center}

\caption{Imaginary part of the dispersion relation $\omega(q)$ at $T=0$
as a function of real $q$ for $\geff^{2}(\muMS=4\mu)=9$ and 36 in
the large-$N_{f}$ limit compared to the HDL result. In contrast to
the imaginary part at zero chemical potential in Figures \ref{fig:dispersionw}
and \ref{fig:dispersionw36}, the imaginary part is zero up to a threshold
$q\geq2\mu-\re\,\omega(q)$ as explained in the text. The constant
HDL line in the right panel is hidden by the exact large-$N_{f}$
lines. The real part of the dispersion relations looks almost identical
to the case of zero chemical potential (Figures \ref{fig:dispersionw}
and \ref{fig:dispersionw36}) and is not plotted separately. \label{fig:dispersionwzeroT}}
\end{figure}
\begin{figure}
\begin{center}\hfill{}\includegraphics[%
  scale=0.8]{
  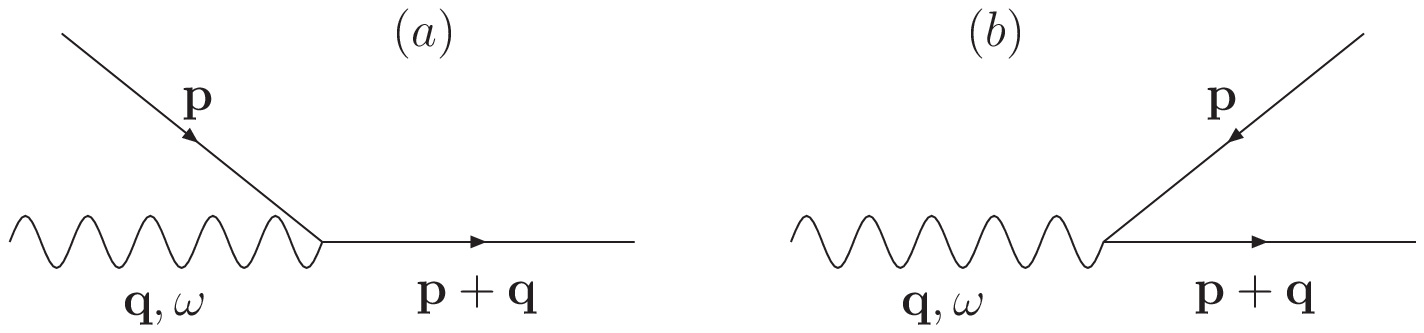}\hfill{}\includegraphics[%
  scale=0.6]{
  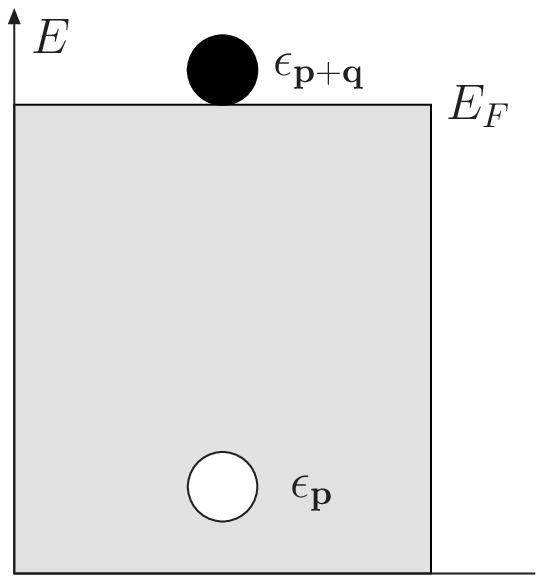}\hfill{}\end{center}

\caption{Diagrammatic processes $(a)$ for Landau damping and $(b)$ for gluon
decay. Only the latter one is allowed kinematically for time-like
gluons with $\omega^{2}>q^{2}$. For non-zero chemical potential and
zero temperature, this process is further restricted to $\omega+q\geq2\mu$
as explained in the text. At the threshold, a quark with Fermi energy
$\mu$ and a collinear anti-quark with energy $\epsilon_{{\bf p}}=(\omega-q)/2$
are produced. 
\label{fig:photondecay}}
\end{figure}

For the propagating modes,
Figure \ref{fig:dispersionwzeroT} shows the imaginary part of the
dispersion relations at zero temperature for the two couplings $\geff^{2}=9$
and 36 at the scale $\muMS=4\mu$. We did not plot the real part of
the function separately, since they are almost indistinguishable from
Figures \ref{fig:dispersionw} and \ref{fig:dispersionw36} at zero
chemical potential. The HDL result for the imaginary part in Fig.~\ref{fig:dispersionwzeroT}
is just zero, and the resulting imaginary part beyond HDL shows a
threshold around the Fermi momentum $q\gtrsim\mu$ which can be explained
as follows.

The process responsible for the imaginary part of the self-energy
and thus for the propagator pole is the pair creation as shown in
Figure \ref{fig:photondecay}. At real momenta $q$ we are looking
at time-like momenta for which the Landau damping process is kinematically
forbidden, even for massive quarks. The only process that could contribute
to the imaginary part of the self-energy is the creation of a quark
with momentum ${\bf p+{\bf q}}$ and an anti-quark with momentum ${\bf p}$.
At finite chemical potential, this process is kinematically restricted
as the Fermi sea is filled up to the Fermi energy $E_{F}=\mu$ and
the newly produced quark must have an energy above, $\epsilon_{{\bf p+q}}\geq\mu$.
The anti-quark can in principle be produced within the Fermi sea,
$\epsilon_{{\bf p}}\geq0$, but the dispersion relation of the gluon
$\omega(q)$ dictates a minimal energy. From energy conservation of
the process $\omega=\epsilon_{{\bf p}}+\epsilon_{{\bf p+q}}$ we obtain
for the angle $\theta$ between ${\bf p}$ and ${\bf q}$\begin{equation}
\cos\theta=\frac{\omega^{2}-q^{2}-2\omega\epsilon_{{\bf p}}}{2pq}\,.\end{equation}
For zero mass, requiring a valid angle $-1\leq\cos\theta\leq1$, we
obtain the kinematical restriction $\omega+q\geq2p\geq\omega-q$.
For a small finite quark mass $m^{2}\leq Q^{2}=\omega^{2}-q^{2}$,
the lower bound for $p$ yields
\begin{equation}
p\geq\frac{1}{2}\left(\omega\gamma(Q)-q\right)\end{equation}
with $\gamma(Q)\equiv\sqrt{1-4m^{2}/Q^{2}}$. From the energy threshold
of the quark $\epsilon_{{\bf p+q}}\geq\mu$ we immediately obtain
\begin{equation}
\omega\gamma(Q)+q\geq2\mu\,,\end{equation}
where $\gamma(Q)=1$ in the massless case. The momentum thresholds
in Figure \ref{fig:dispersionwzeroT} just follow this formula $q\geq2\mu-\re\,\omega(q)$,
scaled by the Debye mass $m_{D}$, where real part of $\omega(q)$
can be read off from Figures \ref{fig:dispersionw} and \ref{fig:dispersionw36}.
It also explains why the momentum threshold for transverse modes is
slightly lower than for the corresponding longitudinal mode, because
$\omega_{T}(q)>\omega_{L}(q)$ for $q>0$. 

\subsection{Friedel oscillations}

For finite chemical potential and comparatively small temperatures,
electrostatic screening at sufficiently large distances is no longer given by 
a simple Yukawa-type potential involving the Debye mass $m_D$.
As is well known in nonrelativistic quantum many-body theory \cite{FetW:Q},
the sharpness of the Fermi surface leads to
an asymptotic behavior of the potential that is governed
by so-called Friedel oscillations, implying a shell-like structure
with alternating screening and overscreening. At $T=0$, the amplitude
of these oscillations decays by a power law and not like an exponential.
As has been shown in Ref.~\cite{Kapusta:1988fi}, a similar modification
of Debye screening occurs in high-density QED and QCD. In the following
we shall investigate this effect quantitatively in the large-$N_f$ theory.

The free energy of two (Abelian) static charges at a distance $r$
is given by the Fourier transform of the electrostatic gauge boson
propagator \cite{Kapusta:1988fi}
\be
V(r)=Q_1 Q_2 \int {d^3k\0(2\pi)^3} e^{ik\cdot r}{1\0k^2+\Pi_L(k_0=0,k)}.
\ee
The deviation from Coulomb law behavior is described by the screening
function
\be\label{sofr}
s(r)={4\pi r V(r)\0 Q_1 Q_2}={1\0\pi}\Im\int_{-\infty}^\infty
{dk\, k\, e^{ikr} \0 k^2+\Pi_L(k_0=0,k)},
\ee
where we have used the fact that $\Pi_L(k_0=0,k)$ is an even function
of $k=\sqrt{k^2}$.
In the HTL approximation, where $\Pi_L(k_0=0,k)=\hat m_D^2$,
one has simply $s(r)=e^{-m_D r}$.

\begin{figure}
\begin{center}\includegraphics[%
  scale=0.66]{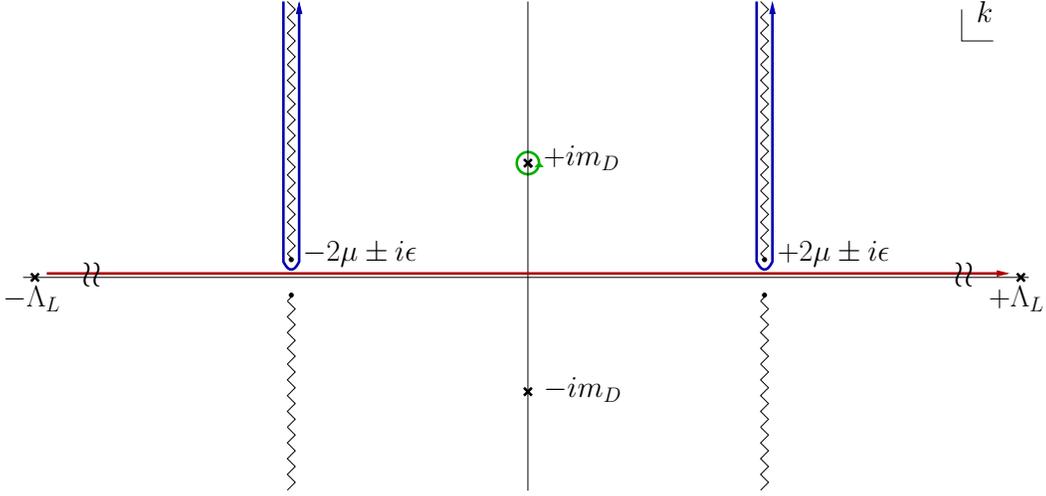}\end{center}
\caption{Analytic structure of the electrostatic propagator
at zero temperature and integration paths for the screening
function (\ref{sofr}). If the Landau poles at real $k=\pm \Lambda_L$
are moved infinitesimally into the lower half-plane, the
integration over the real $k$ axis can be deformed to
encircle the two branch cuts starting at $k=\pm 2\mu + i\epsilon$
and the Debye mass pole at $k=im_D$. 
\label{fig:Friedelpaths}}
\end{figure}

However,
as can be seen from the analytical results in Appendix \ref{appB4},
the full gauge boson propagator 
at zero temperature 
also involves logarithmic branch cuts which
touch the real axis at momenta $k=\pm 2\mu$. Following Refs.~\cite{FetW:Q,Kapusta:1988fi}, we separate these branch cuts infinitesimally
according to $k=\pm 2\mu\pm i\epsilon$ so that the integration
path of Eq.~(\ref{sofr}) is threaded through the four
branch points as shown in Fig.~\ref{fig:Friedelpaths}.
Ignoring (as implicitly done in Ref.~\cite{Kapusta:1988fi})
the Landau poles located at real $k=\pm \Lambda_L \gg \mu$ by moving
them slightly into the lower half-plane, the screening function
receives contributions from the Debye mass pole at $k=im_D$
and two integrations along the cuts starting at $k=\pm 2\mu+i\epsilon$.
Because of the latter, 
the electrical screening at sufficiently large distances 
is dominated by oscillatory behavior with wavelength $\pi/\mu$.
In the ultrarelativistic case
the asymptotic behavior 
has been calculated in Ref.~\cite{Kapusta:1988fi} with the result
\be\label{sofras}
\lim_{r\to\infty} s(r) = 
{\g^2\016\pi^2} \left[1+{\g^2\06\pi^2}\left(\ln{\muMS\04\mu}+{11\06}
\right)
\right]^{-2} {\sin(2\mu r)\0 (\mu r)^3},
\ee
which we have transcribed to
the $\overline{\rm MS}$ scheme that we are using here.
Nonzero temperature
moves the branch cuts away from the real axis, leading to 
an extra exponential suppression factor \cite{Kapusta:1988fi}
$e^{-2\pi Tr}$ so that they cease to govern the asymptotic behavior
for $T\gtrsim m_D/(2\pi)$. We shall therefore concentrate on
the $T=0$ case, where Friedel oscillations should be maximally important.

\begin{figure}
\begin{center}\includegraphics[%
  scale=0.9]{
  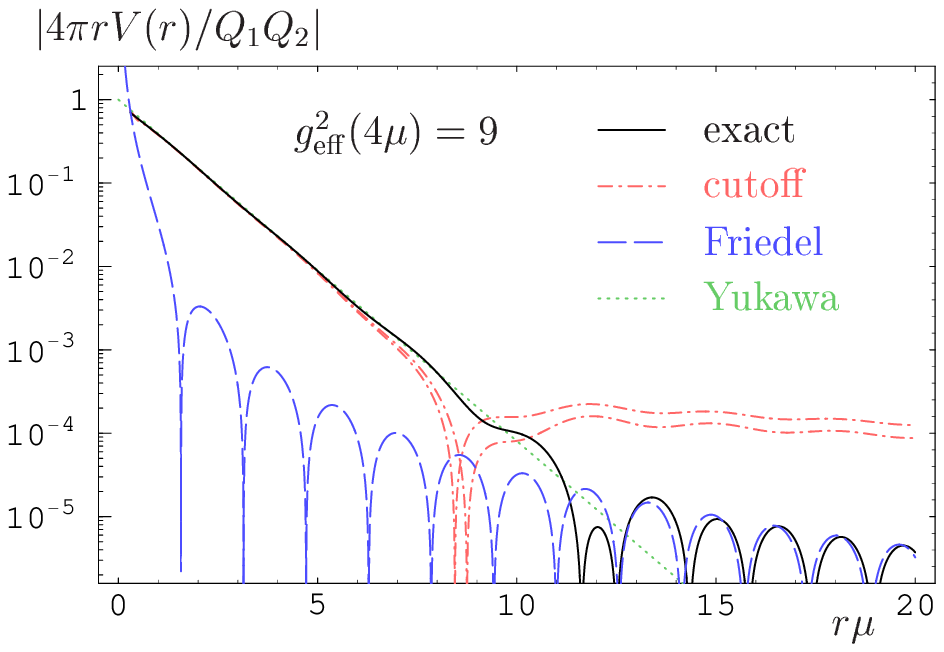}\includegraphics[%
  scale=0.9]{
  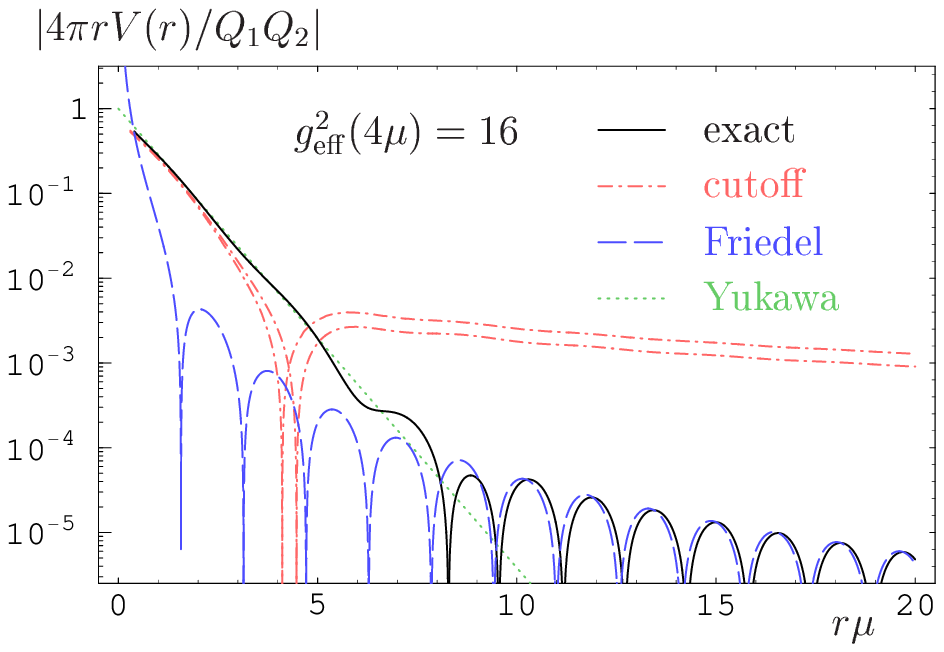}\end{center}
\caption{Electrostatic screening in the large-$N_f$ theory at $T=0$ for
effective coupling $\g^2(\muMS\!=\!4\mu)=9$ and 16. The full line
labelled ``exact'' shows the numerical result obtained by circumventing
the Landau pole in the upper half-plane (see Fig.~\ref{fig:Friedelpaths});
the lines labelled ``cutoff'' correspond to cutoffs at the
last zero of the integrand in (\ref{sofr}) below $|k|= a\Lambda_L$
with $a=1/2\ldots 1/\sqrt2$. The dotted line labelled ``Yukawa''
corresponds to an exponential with exact large-$N_f$ screening mass $m_D(\g)$,
the long-dashed line labelled ``Friedel'' represents the asymptotic
formula (\ref{sofras}).
\label{fig:Friedel}}
\end{figure}

With regard to potential implications to high-density QED or QCD 
\cite{Kapusta:1988fi}, it is of interest to know whether
deviations from standard Debye screening occur already at moderate
distances before the screening function has become negligibly small.
This requires a numerical evaluation of the integral in
Eq.~(\ref{sofr}) and the results for $\g^2=9$ and $\g^2=16$ are
shown in Fig.~\ref{fig:Friedel}. The dotted line shows the
contribution from the Debye mass pole with values $m_D= 0.94269\mu$
and $m_D=1.24705\mu$, respectively; the long-dashed lines labelled
``Friedel'' correspond
to the asymptotic formula (\ref{sofras}). The full large-$N_f$
result which excludes contributions from the Landau poles
by moving them infinitesimally into the lower half-plane is
given by the line labelled ``exact''. As one can see, significant deviations
from Debye screening appear only at such large distances that
the screening function $s(r)$ is of the order $10^{-4}$ or below
for $\g^2=9$ and $\g^2=16$. At these couplings, the Landau pole
is at scales of the order of $10^4\mu$ and $3\times 10^2\mu$, respectively.
If instead of circumventing the Landau pole we cut off the
momentum integration at $a\Lambda_L$ with $a$ varied between
$1/2$ and $1/\sqrt2$, this affects the screening function
already considerably above the scale where Friedel oscillations
occur, as shown by the dash-dotted lines in Fig.~\ref{fig:Friedel}.%
\footnote{A cutoff independent of $r$ actually would give rapid oscillations
at large distances. In Fig.~\ref{fig:Friedel} we have instead
pinned the cutoff to the last zero of the oscillatory integrand
below $a\Lambda_L$.}
In real QCD, the one-loop Dyson-resummed gluon propagator
also involves vacuum contributions from the gluon loop which
gives asymptotic freedom and eliminates the Landau pole.\footnote{It
also introduces a branch cut lying on the imaginary axis, which is
however not stable against higher-order contributions.} Because of this,
the results obtained by simply circumventing the Landau pole in the
large-$N_f$ theory are probably a reasonable model of screening
in high-density QCD at low temperature. Taken as such, these
results suggest that Friedel oscillations occur only at distances
where the screened potential is already very small.
At low but nonzero temperature, the importance of 
Friedel oscillations is further reduced by
an extra exponential suppression factor \cite{Kapusta:1988fi}
$e^{-2\pi Tr}$ until they disappear completely
for $T\gtrsim m_D/(2\pi)$.

In ultrarelativistic (massless)
QED, where $\g^2\approx 0.092$, Friedel oscillations
are much more suppressed than in the examples above. 
Ordinary Debye screening dominates
out to distances $r\sim 250 \mu^{-1}$, where the screening function
$s \lesssim 10^{-10}$. However, Friedel oscillations in massive QED
are less strongly suppressed -- they then come with an extra factor of $m r$
in the screening function \cite{FetW:Q,Kapusta:1988fi}.
Indeed, Friedel oscillations have been seen experimentally 
in the nonrelativistic electron gas.

%



\section{Conclusions}

In this paper we studied the spectral functions and dispersion relations
of the gauge boson propagator in the large-$N_{f}$ limit. 
This allows us to
investigate the properties of this propagator not
only in the weak coupling regime around $\geff^{2}(\pi T)\lesssim4$,
but also in the strong coupling regime $4\lesssim\geff^{2}(\pi T)\lesssim36$,
where strict perturbation theory fails but the influence of the Landau
pole is still negligible. For larger values of the coupling, the usual
spectral function sum rules have to be corrected due to the presence
of the Landau pole.

Numerically we calculated the Debye mass, the plasma frequency, dispersion
relations, and spectral functions at finite temperature and compared
those to the analytically known HTL results. The poles of the propagators
at large $N_{f}$ are renormalization scale independent. Depending
on their vicinity to the real energy axis, they appear as peaks of
finite width in the spectral functions. These poles do not lie on
the physical sheet of the propagator, so a proper analytic continuation
of the self-energies into the neighbouring unphysical sheet is necessary.
In addition to moving the branch cuts of the logarithms involved as
it is sufficient for the vacuum piece of the self-energy, one also
has to adjust the integration path of a numerical integration appearing
within the thermal piece of the self-energy in order to avoid a crossing
with logarithmic singularities or branch cuts.


In the large-$N_f$ limit, the full Debye mass is remarkably close
to the HTL value for renormalization scale $\muMS\approx \mu T$
even up to extremely large couplings.
The plasma frequency, and to lesser degree the frequency of
the propagating modes,
acquire an imaginary part which is caused by
the vacuum process of quark-antiquark pair creation neglected
in the HTL approximation. The peaks in
the spectral functions get broader with increasing coupling and their
frequency relative to the HTL value decreases. 
{\sf Overall, there do not seem to be significant qualitative changes
in the gauge boson propagator as one increases the coupling
from small to rather large values.
By contrast, the interaction pressure of the large-$N_f$ theory
exhibits nonmonotonic behavior. As will be shown in a subsequent
paper \cite{Birr}, the latter can be better understood in
terms of the entropy. While the nontrivial content of the
large-$N_f$ pressure can
be formulated entirely in terms of the gauge boson propagator, 
a quasiparticle analysis of the entropy shows that
the bosonic contribution is unsurprising, 
whereas the fermionic contributions involve
next-to-leading order results for the asymptotic quark mass
which accounts for the nonmonotonic behavior at large coupling.

We have also investigated quantitatively 
the effect of asymptotic Friedel oscillations in electrostatic screening
at zero temperature and high chemical potential with negligible
fermion mass.
While Friedel oscillations are a truly tiny effect
in ultrarelativistic (massless) QED, their importance
increases with effective coupling strength. However,
the potential of static charges remains 
dominated by Yukawa-like Debye screening
up to distances where the screening function has dropped by several orders
of magnitude and where the effect of a cutoff to remove the Landau
pole would be of comparable or greater importance.
}

\subsection*{Acknowledgments}

The numerical integration paths have mostly been drawn with the packages
\noun{Axodraw} \cite{Vermaseren:1994je} and \noun{Jaxodraw} \cite{Binosi:2003yf}.

\appendix

\section{Bosonic one-loop self-energy\label{sec:Pi}}

\subsection{Full one-loop result at zero mass}

The vacuum part of the bosonic self energy in the Feynman gauge (covariant
gauge with gauge parameter $\alpha=1$) for $N_{f}$ flavors and $N_{c}$
colors, coupling $\alpha_{s}=g^{2}/(4\pi)$, and the renormalization
scale $\muMS$ in the modified minimal subtraction ($\overline{{\rm MS}}$)
scheme is given by\begin{equation}
\Pi_{{\rm \vac}}(Q)=\frac{g^{2}}{16\pi^{2}}Q^{2}\left\{ \frac{2N_{f}-5N_{c}}{3}\left[\log\left(\frac{-Q^{2}}{\muMS^{2}}\right)-\frac{5}{3}\right]+\frac{2}{3}N_{c}\right\} .\label{Pivac}\end{equation}

The thermal part of the bosonic self energy
reads \cite{Weldon:1982aq}
\begin{eqnarray}
\Pi_{{\rm L,th}}(q_{0},q) & = & g^{2}\left(N_{f}H_{f}+N_{c}H_{b}\right),\label{PiL}\\
\Pi_{{\rm T,th}}(q_{0},q) & = & \frac{g^{2}}{2}\left(-\frac{q^{2}-q_{0}^{2}}{q^{2}}\left(N_{f}H_{f}+N_{c}H_{b}\right)+\left(N_{f}G_{f}+N_{c}G_{b}\right)\right),\label{PiT}\end{eqnarray}
for $N_{f}$ flavors and $N_{c}$colors, where the fermionic functions
$G_{f}$ and $H_{f}$ can be written as one-dimensional integrals

\begin{eqnarray}
G_{f} & = & \frac{1}{2\pi^{2}}\int_{0}^{\infty}dk\, n_{f}(k)\left(4k-\frac{q^{2}-q_{0}^{2}}{2q}L_{1}\right),\label{Gf}\\
H_{f} & = & \frac{1}{2\pi^{2}}\int_{0}^{\infty}dk\, n_{f}(k)\left(2k-\frac{q^{2}-q_{0}^{2}-4k^{2}}{4q}L_{1}-q_{0}kL_{2}\right),\label{Hf}\end{eqnarray}
and similarly the bosonic functions $G_{b}$ and $H_{b}$
\begin{eqnarray}
G_{b} & = & \frac{1}{2\pi^{2}}\int_{0}^{\infty}dk\, n_{b}(k)\left(4k-\frac{5}{4}\frac{q^{2}-q_{0}^{2}}{q}L_{1}\right),\label{Gb}\\
H_{b} & = & \frac{1}{2\pi^{2}}\int_{0}^{\infty}dk\, n_{b}(k)\left(2k-\frac{2q^{2}-q_{0}^{2}-4k^{2}}{4q}L_{1}-q_{0}kL_{2}\right),\label{Hb}\end{eqnarray}
with\begin{eqnarray}
L_{1} & = & \log\left(\frac{2k+q-q_{0}}{2k-q-q_{0}}\right)-\log\left(\frac{2k-q+q_{0}}{2k+q+q_{0}}\right),\label{L1}\\
L_{2} & = & \log\left(\frac{2k+q-q_{0}}{2k-q-q_{0}}\right)-2\log\left(\frac{-q+q_{0}}{q+q_{0}}\right)+\log\left(\frac{2k-q+q_{0}}{2k+q+q_{0}}\right),\,\,\label{L2}\end{eqnarray}
and the bosonic and fermionic distribution functions at temperature
$T$ and chemical potential $\mu$\begin{eqnarray}
n_{b}(k) & = & \frac{1}{e^{k/T}-1},\label{nb}\\
n_{f}(k) & = & \frac{1}{2}\left(\frac{1}{e^{(k-\mu)/T}+1}+\frac{1}{e^{(k+\mu)/T}+1}\right).\label{nf}\end{eqnarray}
These expressions are valid for all complex $q_{0}$ in a rotation
from Euclidean space $q_{0}=i\omega$ to Minkowski space $q_{0}=\omega+i\epsilon$
with $\epsilon>0$. For the analytic continuation into the region
with $\epsilon<0$ the expressions $L_{1}$ and $L_{2}$ have to be
changed as described below.

{\sf Apart from their HTL/HDL content, the bosonic contributions 
(\ref{Gb}) and (\ref{Hb}) are gauge dependent (see Refs.~\cite{Kajantie:1985xx,Heinz:1987kz} for one-loop results in gauges other than Feynman).}

\subsection{HTL and HDL\label{sec:HTL}}

The corresponding HTL or HDL expressions can be extracted by demanding
that the external momenta $q_{0}$ and $q$ be small compared to
temperature $T$ and/or chemical potential $\mu$ (that is to take
the leading contribution in a $1/k$ expansion of the integrand, since
the main contribution is expected to come from large loop momenta
$k$). The relevant physical scale is then given by the HTL Debye
mass (HTL quantities are marked by a $\hat{\:}$-hat)\begin{equation}
\hat{m}_{D}^{2}=g^{2}\left\{ \frac{2N_{c}+N_{f}}{6}T^{2}+N_{f}\frac{\mu^{2}}{2\pi^{2}}\right\} .\label{eq:mD}\end{equation}
In this limit, the equations (\ref{Gf}) to (\ref{Hb}) reduce to
\begin{eqnarray}
\hat{G}_{f} & = & \frac{T^{2}}{6}+\frac{\mu^{2}}{2\pi^{2}},\label{GfHTL}\\
\hat{H}_{f} & = & \left(\frac{T^{2}}{6}+\frac{\mu^{2}}{2\pi^{2}}\right)\left\{ 1+\frac{q_{0}}{2q}\log\left(\frac{-q+q_{0}}{q+q_{0}}\right)\right\} ,\label{HfHTL}\\
\hat{G}_{b} & = & \frac{T^{2}}{3},\label{GbHTL}\\
\hat{H}_{b} & = & \frac{T^{2}}{3}\left\{ 1+\frac{q_{0}}{2q}\log\left(\frac{-q+q_{0}}{q+q_{0}}\right)\right\} .\label{HbHTL}\end{eqnarray}
In the zero temperature limit $\hat{G}_{b}$ and $\hat{H}_{b}$ vanish.

\section{Analytic continuation of the self-energy\label{sec:analyticcontinuation}}


\subsection{Integration in the complex $k$ plane}

\begin{figure}
\begin{center}\includegraphics[%
  scale=0.7]{
  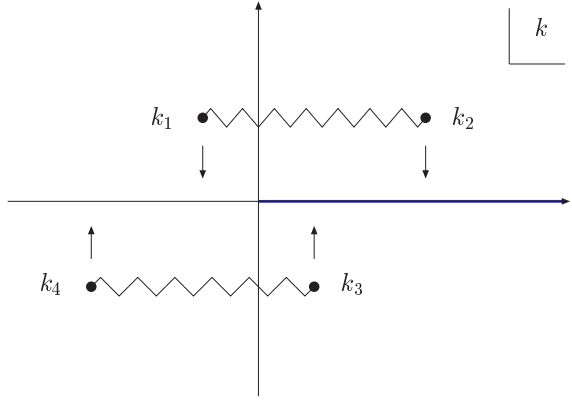}\end{center}

\caption{Branch cut and singularity structure of the initial integrand for
$G_{f}$, Eq.~(\ref{Gf}), for complex $k$. The singular points
$k_{1}$ to $k_{4}$ correspond to solutions of $2k\pm q\pm q_{0}=0$
from the logarithmic argument in $L_{1}$, Eq.~(\ref{L1}), which
are pairwise connected by logarithmic branch cuts. For $q_{0}=\omega+i\epsilon$,
all four singularities will simultaneously cross the real $k$
axis as $\epsilon$ changes from positive to negative sign. \label{fig:polestructure1}}
\end{figure}

The $k-$integration in the self-energy expressions for $G_{f,b}$
and $H_{f,b}$ in equations (\ref{Gf}) to (\ref{Hb}) runs over real
$k$ only. One might assume that for the analytic continuation of
those functions it is sufficient to take care of the logarithms appearing
in $L_{1}$ and $L_{2}$ and leave the real $k$ integration untouched.
It turns out that this simple approach fails. For continuing the functions
properly in the complex domain, we should think of the $k-$integration
as an analytic integration in the complex $k$ plane. $L_{1}$ and
$L_{2}$ provide singular points and branch cuts in the complex $k$
plane that move around, as $\epsilon$ varies. As long as the logarithmic
singularities or branch cuts never touch the integration line, the
resulting integral is an analytic, complex function. We will demonstrate
the necessary procedure for the branch cut structure of $L_{1}$ of
$G_{f}$, but the following considerations can be readily applied
to $L_{2}$ and therefore to $G_{b}$, $H_{f}$, and $H_{b}$. 

Figure \ref{fig:polestructure1} shows the singularity and branch
cut structure for the integrand of $G_{f}$, Eq.~(\ref{Gf}), which
is determined for complex $k$ by the logarithms in $L_{1}$, Eq.~(\ref{L1}).
For $q_{0}=\omega+i\epsilon$, the logarithmic singularities lie at
\begin{eqnarray}
k_{1} & = & \frac{1}{2}(-q+\omega+i\epsilon),\nonumber \\
k_{2} & = & \frac{1}{2}(q+\omega+i\epsilon),\nonumber \\
k_{3} & = & \frac{1}{2}(q-\omega-i\epsilon)\,=\,-k_{1},\nonumber \\
k_{4} & = & \frac{1}{2}(-q-\omega-i\epsilon)\,=\,-k_{2},\label{k1to4}\end{eqnarray}
so with $\epsilon>0$ and real $q$ and $\omega$ we find that $k_{1}$
and $k_{2}$ lie initially above the real $k$ axis, while $k_{3}$
and $k_{4}$ lie below. The branch cut structure shown in Figure \ref{fig:polestructure1}
corresponds to the initial formula for $L_{1}$ from Eq.~(\ref{L1}):
$k_{1}$ and $k_{2}$ are connected by a logarithmic branch cut, as
are $k_{3}$ and $k_{4}$. The problem is evident as we try to analytically
extend the integration from the region $\epsilon>0$ to $\epsilon<0$:
All four singularities will pass the real axis at the same time, as
do the corresponding branch cuts that penetrate each other. How can
we choose an integration path that avoids contact with these singularities
and branch cuts? The solution lies in the fact that we can move around
the branch cuts. 

\begin{figure}
\begin{center}\includegraphics[%
  scale=0.7]{
  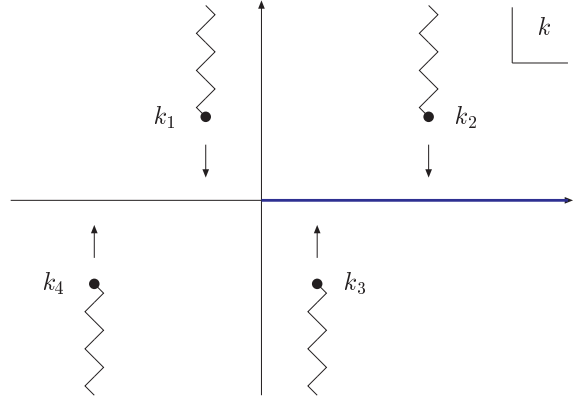}\end{center}

\caption{First we break up the branch cuts according to Eq.~(\ref{L1opened}).\label{fig:polestructure2}}
\end{figure}
\begin{figure}
\begin{center}\includegraphics[%
  scale=0.7]{
  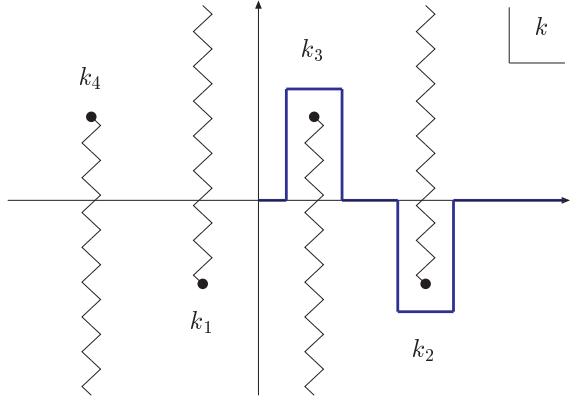}\end{center}

\caption{In this way, we can deform the integration path when the singularities
and branch cuts move across the real $k$ axis. Note that we could have
used this integration path already in figure \ref{fig:polestructure2}.
In that case it is obvious that moving the singularities across the
real $k$ axis would not be different from moving the singularities
around anywhere else in the complex $k$ plane. \label{fig:polestructure3}}
\end{figure}

First we break up the branch cuts from the logarithmic expressions
in $L_{1}$ and $L_{2}$ as shown in Figure \ref{fig:polestructure2}.
Using the logarithmic functions $\log^{\uparrow}(z)$ and $\log^{\downarrow}(z)$
from Eqs.~(\ref{logup}) and (\ref{logdown}) with branch cuts along
the positive or negative imaginary axis, we can rewrite equation (\ref{L1})
as

\begin{eqnarray}
L_{1}^{{\rm break-up}} & = & \log^{\uparrow}(2k+q-q_{0})-\log^{\uparrow}(2k-q-q_{0})\nonumber \\
 &  & -\log^{\downarrow}(2k-q+q_{0})+\log^{\downarrow}(2k+q+q_{0})\,,\label{L1opened}\end{eqnarray}
which agrees with $L_{1}$ on the real $k$ axis, and similarly for
$L_{2}$ \begin{eqnarray}
L_{2}^{{\rm break-up}} & = & \log^{\uparrow}(2k+q-q_{0})-\log^{\uparrow}(2k-q-q_{0})\nonumber \\
 &  & -2\left(\log^{\downarrow}(-q+q_{0})-\log^{\downarrow}(q+q_{0})\right)\nonumber \\
 &  & +\log^{\downarrow}(2k-q+q_{0})-\log^{\downarrow}(2k+q+q_{0})\,.\label{L2opened}\end{eqnarray}
The branch cuts in the two logarithmic expressions of $L_{2}$ independent
of $k$ (second line) are chosen such that the functions continue
analytically for real $q$ and $\omega$ when $\epsilon$ changes
its sign from positive to negative. Using $L_{1}^{{\rm break-up}}$
gives the branch cut structure as shown in Figure \ref{fig:polestructure2}.
It is now clear how the path should be deformed from $\epsilon>0$
in Fig.~\ref{fig:polestructure2} to $\epsilon<0$ in Fig.~\ref{fig:polestructure3}.
To further justify this procedure, let us note that one could have
used the new integration path from Fig.~\ref{fig:polestructure3}
already in Fig.~\ref{fig:polestructure2}, where it would not have
changed the result of the integration, as the result of the complex
integration is of course path independent. Keeping the same S-shaped
integration path in both figures, it is clear that moving a logarithmic
singularity across the real $k$ axis is not different from moving
it around anywhere else in the complex $k$ plane, provided no
singularities are crossed.

Practically, to change from positive to negative $\epsilon$, we could
either literally choose a numerical integration contour similar to
Figure \ref{fig:polestructure3}, or we could basically stick to the
original integration along the real $k$ axis, taking correcting branch
cut contributions from the real axis to, say, $k_{2}$ and $k_{4}$
into account. At first sight, the latter strategy seems insofar appealing,
as the branch-cut contributions can be calculated analytically without
the necessity of numerical integration. This strategy is briefly discussed
below in Appendix \ref{sec:alternativeways}. But since the integration
parts along the real $k$ axis have to be integrated numerically anyway,
it turns out to be easier to choose an appropriate integration path
for numerical integration in the complex $k$ plane which avoids all
branch cuts in a safe distance.

\subsection{Choosing a complex integration path}

\begin{figure}
\begin{center}\includegraphics[%
  scale=0.7]{
  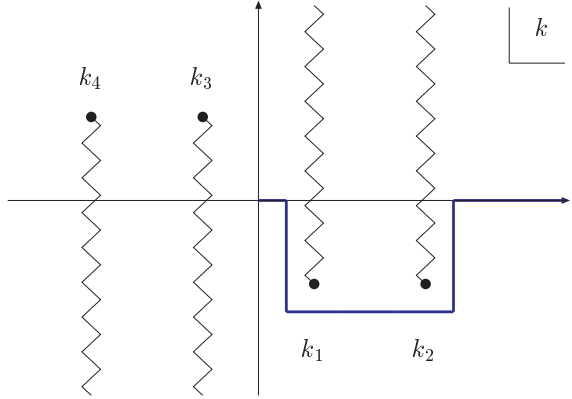}\end{center}

\caption{If $\re\, k_{1}>0$ and $\re\, k_{3}<0$ then the path would look
like this. In Figures \ref{fig:polestructure1} to \ref{fig:polestructure3}
we had assumed $\re\, k_{3}>0$ and $\re\, k_{1}<0$. \label{fig:polestructure4}}
\end{figure}
We shall show how to choose an integration path in the complex $k$
plane that circumvents the branch cuts. In choosing the path, we only
need to care about logarithmic singularities $k_{i}$ with positive
real part as only those will cross the initial integration range $k\in[0,\infty)$.
Assuming that $\re\, q>0$ and real $\omega>0$ in $q_{0}=\omega+i\epsilon$,
we see that in (\ref{k1to4}) the singularity $\re\, k_{2}>0$ always
lies in the right half of the complex plane while $\re\, k_{4}<0$
lies on the left half. Also $k_{1}$ and $k_{3}$ lie on opposite
sides of the imaginary $k$-axis, so there are two distinct cases
for the path topology: Either $\re\, k_{1}>0$ as in figure \ref{fig:polestructure4},
or $\re\, k_{3}>0$ as in figure \ref{fig:polestructure3}. In the
latter case, $\re\, k_{3}$ is always smaller than $\re\, k_{2}$,
but in the former case $\re\, k_{1}$ could be either smaller or larger
than $\re\, k_{2}$. 

\begin{figure}
\begin{center}\includegraphics[%
  scale=0.7]{
  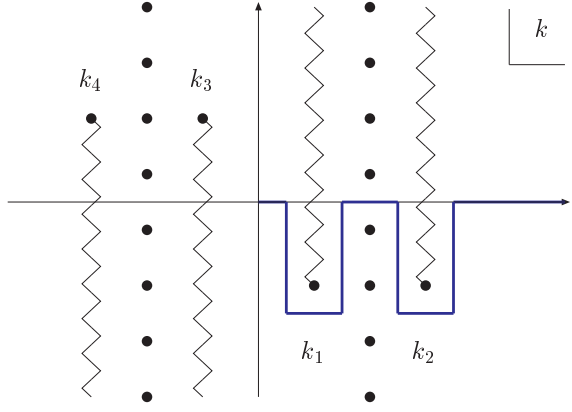}\end{center}

\caption{At finite chemical potential $\mu$, the fermionic distribution function
(\ref{nf}) provides additional poles at the Matsubara frequencies
$k=\pm\mu+(2n+1)i\pi T$ ($n\in\mathbb{Z}$) which have to be avoided.
\label{fig:polestructure5}}
\end{figure}

At finite chemical potential the fermionic distribution function $n_{f}(k)$
from Eq. (\ref{nf}) provides additional poles at $k=\pm\mu+(2n+1)i\pi T$
($n\in\mathbb{Z}$). Figure \ref{fig:polestructure5} shows the particular
case $0<\re\, k_{1}<\mu<\re\, k_{2}$, but the Matsubara frequencies
can also lie to the left of $\re\, k_{1}$ or to the right of $\re\, k_{2}$.
As the initial path for the $k$-integration (starting from $\epsilon>0$)
lies along the real $k$ axis between the poles $k=\mu\pm i\pi T$,
also after the analytic continuation from $\epsilon>0$ to $\epsilon<0$
the integration path has to stay between these two lowest Matsubara
frequency poles. Similarly, one has to take care of the Matsubara
frequency poles in the case $\re\, k_{3}>0$ as in figure \ref{fig:polestructure3}.

There is a lot of freedom in choosing an integration path that fulfills
all of these constraints and correctly avoids the problematic regions
in the complex plane. Our implementation of the complex integration
path follows essentially those suggested in Figures \ref{fig:polestructure3},
\ref{fig:polestructure4}, and \ref{fig:polestructure5}, where we
vary the minimal distances to the branch cuts and singularities for
testing purposes. One can readily check that the final result is independent
of the exact implementation of a correct integration path.

The same integration path can be used for calculating the derivative
with respect to $q_{0}$, by forming the derivative of the integrand
first. The derivative of the self-energy functions can be used in
specifying the Jacobian for a numerical root search. The search for
the Debye mass requires complex $q$. Assuming $\re\, q\geq0$ and
$\im\, q\geq0$ this implies straightforward modifications to the
integration paths: As the real part of $q(q_{0})$ vanishes in the
limit $q_{0}\rightarrow0$, basically $k_{1}$ and $k_{2}$ from Eqs.~(\ref{k1to4})
come to lie vertically aligned on top of each other in Fig.~\ref{fig:polestructure4}
instead of horizontally aligned side by side.

\subsection{Alternative ways of calculating the analytic continuation\label{sec:alternativeways}}

One possibility (which we did not follow in our numerical evaluation)
is to analytically calculate the contribution caused by a branch cut
that crosses our original integration line. Integrating a function
$f(z)\log(z)$ (where $f(z)$ is analytic) around a part of the branch
cut is equivalent to integrating $f(z)2\pi i$ along the path of the
branch cut, e.g. with $a<b<0$ we can write\begin{equation}
\left[\int_{a-i\epsilon}^{b-i\epsilon}+\int_{b+i\epsilon}^{a+i\epsilon}\right]f(z)\log(z)dz=\int_{b}^{a}f(z)2\pi i\, dz\,,\end{equation}
as the real parts of $\log(z)=\log|z|+i\arg z$ along the two contour
parts cancel each other, while the net contribution comes from integrating
$\pm i\pi$ along the contour.

For $G_{f,b}$ we basically have to integrate $n_{f,b}(k)\times{\rm polynomial}(k)\times2\pi i$.
This can be done analytically. Here are some examples for $n_{b}$\begin{eqnarray}
\int n_{b}(k)dk & = & T\log\left(e^{k/T}-1\right)-k,\\
\int kn_{b}(k)dk & = & T^{2}\Li_{2}\left(e^{k/T}\right)+kT\log\left(1-e^{k/T}\right)-\frac{k^{2}}{2},\\
\int k^{2}n_{b}(k)dk & = & -2T^{3}\Li_{3}\left(e^{k/T}\right)+2kT^{2}\Li_{2}\left(e^{k/T}\right)\nonumber \\
 &  & +k^{2}T\log\left(1-e^{k/T}\right)-\frac{k^{3}}{3},\end{eqnarray}
where $\Li_{n}(z)$ is the polylogarithm function. Similar expressions
can be derived for $n_{f}$. These functions have branch cuts for
$\re(k)>0$ which would have to be corrected, e.g. \begin{equation}
\int n_{b}(k)dk=T\log\left(e^{k/T}-1\right)-k-2\pi i\left\lceil \frac{\im(k)}{2\pi T}-\frac{1}{2}\right\rceil \,,\end{equation}
where the ceiling function $\left\lceil x\right\rceil $ means the
smallest integer $\geq x$. Similar adjustments would be necessary
for the other relations above. Since these adjustments are error-prone,
and we have to integrate a large part of the result numerically anyway,
it turns out to be more straightforward to choose a complex integration
path and integrate everything numerically, as described in the previous
section.

Another possibility to obtain the analytically continued propagator
is to exploit the relation between propagator and spectral function,
Eq.~(\ref{eq:spectralsumrule}). Apart from the fact that this would
introduce another numerical integration over $q_{0}$, this relation
is only reliable for small couplings where the influence of the Landau
pole can be neglected. For larger couplings, the correction term of
Eq.~(\ref{eq:spectralsumrulecorrected}) would have to be taken into
account.

Yet another way to obtain the analytically continued retarded propagator
is given by adding a correction term to the Feynman propagator (a
detailed description of this procedure can be found in Appendix B
of Ref.~\cite{Kitazawa:2005vr})\begin{equation}
G^{R}(\omega-i\epsilon,q)=G(\omega-i\epsilon)+2iI(\omega-i\epsilon,q)\,.\end{equation}
Along the discontinuity, i.e.~for real $\omega$, the correction
term coincides with the spectral function from Eq.~(\ref{eq:spectraldefinitionintro}),
$I(\omega,q)=\rho(\omega,q)$, but different from the spectral function,
the function $I(\omega-i\epsilon,q)$ has to be an analytic function.
By the uniqueness theorem of the analytic continuation, one obtains
the unique retarded Greens function $G^{R}(\omega-i\epsilon,q)$.
The tricky point of this approach is to find the correct analytic
continuation of the spectral function $\rho(q_{0},q)$. 
We can not readily apply
this method to the large-$N_{f}$ limit, as the spectral function
is, just like the propagator itself, only obtained numerically. The
proper analytic continuation of the function $I(q_{0},q)$ would be
just as tough as the analytic continuation of $G(q_{0},q)$ itself.

\subsection{Analytic result at zero temperature and finite chemical potential}\label{appB4}

The fermionic distribution function (\ref{nf}) can be expanded for
small $T$ as\begin{eqnarray}
n_{f}(k,T,\mu) & = & \frac{1}{2}\left(\theta(-(k-\mu))+\theta(-(k+\mu))\right)\nonumber \\
 &  & +T^{2}\frac{\pi^{2}}{12}\left(\delta'(k-\mu)+\delta'(k+\mu)\right)+O(T^{4})\label{nf_expansion}\end{eqnarray}
where $\delta'(k-\mu)$ is to be understood in the
sense of distributions
\begin{equation}
\int_{-\infty}^{\infty}\delta'(k-\mu)f(k)dk=-f'(\mu).\label{deltaprimeintegral}\end{equation}
In this limit, the integrations in (\ref{Gf}) to (\ref{Hb}) can
be performed analytically. The analytic continuation is straightforward
with the tools presented here. The result at zero temperature \cite{Ipp:2004qt}
can be written using \begin{eqnarray}
R(q_{0},q) & = & (2\mu+q_{0}+q)\log(2\mu+q_{0}+q)-(q_{0}+q)\log(q_{0}+q),\\
S(q_{0},q) & = & (2\mu+q_{0}-2q)(2\mu+q_{0}+q)^{2}\log(2\mu+q_{0}+q)\\
 &  & -(q_{0}-2q)(q_{0}+q)^{2}\log(q_{0}+q).\nonumber \end{eqnarray}
We use the following abbreviation to keep a lot of terms in dense
notation\begin{equation}
R_{\pm}^{\pm}(q_{0},q)\equiv R(q_{0},q)-R(q_{0},-q)+R(-q_{0},q)-R(-q_{0},-q)\end{equation}
and similarly for $S_{\pm}^{\pm}$ (just replace $R$ by $S$). The
functions $G_{f}^{(0)}$ and $H_{f}^{(0)}$ (the upper index {}``$(0)$''
denotes the $T=0$ limit) can then be written as\begin{eqnarray}
G_{f}^{(0)}(q_{0},q) & = & \frac{1}{2\pi^{2}}\left(\mu^{2}+\frac{q_{0}^{2}-q^{2}}{8q}R_{\pm}^{\pm}(q_{0},q)\right),\\
H_{f}^{(0)}(q_{0},q) & = & \frac{1}{96\pi^{2}}\left(32\mu^{2}+\frac{1}{q}\left(-24\mu^{2}q_{0}\left(\log(q_{0}+q)-\log(q_{0}-q)\right)+S_{\pm}^{\pm}(q_{0},q)\right)\right).\end{eqnarray}
In order to correctly analytically continue these formulae into the
region of $\epsilon<0$, one needs to replace all logarithms that
contain an initially positive (negative) $q_{0}$ to $\log^{\downarrow}$
($\log^{\uparrow}$) from equations (\ref{logup}) and (\ref{logdown}).
Defining $R^{\downarrow(\uparrow)}(q_{0},q):=[R(q_{0},q)$ with $\log\rightarrow\log^{\downarrow(\uparrow)}${]}
and similarly for $S$, we can write 

\begin{equation}
R_{\pm\uparrow}^{\pm\downarrow}(q_{0},q)\equiv R^{\downarrow}(q_{0},q)-R^{\downarrow}(q_{0},-q)+R^{\uparrow}(-q_{0},q)-R^{\uparrow}(-q_{0},-q)\end{equation}
and the very same for $S$. The analytically continued functions $G_{f}^{(0)}$
and $H_{f}^{(0)}$ can then be written as\begin{eqnarray}
G_{f}^{(0)}(q_{0},q) & = & \frac{1}{2\pi^{2}}\left(\mu^{2}+\frac{q_{0}^{2}-q^{2}}{8q}R_{\pm\uparrow}^{\pm\downarrow}(q_{0},q)\right),\\
H_{f}^{(0)}(q_{0},q) & = & \frac{1}{96\pi^{2}}\left(32\mu^{2}+\frac{1}{q}\left(-24\mu^{2}q_{0}\left(\log^{\downarrow}(q_{0}+q)-\log^{\downarrow}(q_{0}-q)\right)+S_{\pm\uparrow}^{\pm\downarrow}(q_{0},q)\right)\right).\end{eqnarray}

The next term in the small $T$ expansion of equation (\ref{nf_expansion})
is easily calculated using (\ref{deltaprimeintegral}) and can be
written as (the upper index {}``$(2)$'' denotes the second derivative
w.r.t.~$T$ at $T=0$)\begin{eqnarray}
G_{f}^{(2)}(q_{0},q) & = & \frac{1}{24}\frac{16\mu^{2}(4\mu^{2}-3q_{0}^{2}-q^{2})}{16\mu^{4}-8\mu^{2}(q_{0}^{2}+q^{2})+(q_{0}^{2}-q^{2})^{2}},\\
H_{f}^{(2)}(q_{0},q) & = & \frac{1}{24q}\left[(2\mu+q_{0})\left(\log(2\mu+q+q_{0})-\log(2\mu-q+q_{0})\right)\right.\nonumber \\
 &  & -2q_{0}\left(\log(q_{0}+q)-\log(q_{0}-q)\right)\nonumber \\
 &  & \left.+(2\mu-q_{0})\left(\log(2\mu+q-q_{0})-\log(2\mu-q-q_{0})\right)\right].\end{eqnarray}
As $G_{f}^{(2)}$ is regular in the whole $q_{0}$ complex plane,
we only have to take care of the proper analytic continuation of $H_{f}^{(2)}$:\begin{eqnarray}
H_{f}^{(2)}(q_{0},q) & = & \frac{1}{24q}\left[(2\mu+q_{0})\left(\log^{\downarrow}(2\mu+q+q_{0})-\log^{\downarrow}(2\mu-q+q_{0})\right)\right.\nonumber \\
 &  & -2q_{0}\left(\log^{\downarrow}(q_{0}+q)-\log^{\downarrow}(q_{0}-q)\right)\nonumber \\
 &  & \left.+(2\mu-q_{0})\left(\log^{\uparrow}(2\mu+q-q_{0})-\log^{\uparrow}(2\mu-q-q_{0})\right)\right].\end{eqnarray}
The analytically continued expansion at small $T$ and finite chemical
potential is then given as\begin{eqnarray*}
G_{f}(q_{0},q) & = & G_{f}^{(0)}(q_{0},q)+T^{2}G_{f}^{(2)}(q_{0},q)+O(T^{4}),\\
H_{f}(q_{0},q) & = & H_{f}^{(0)}(q_{0},q)+T^{2}H_{f}^{(2)}(q_{0},q)+O(T^{4}).\end{eqnarray*}


\end{document}